\pdfoutput=1
\documentclass[traditabstract]{aa} 
\usepackage{graphicx}
\usepackage{txfonts}
\usepackage{natbib}
\usepackage{color}
\bibpunct{(}{)}{;}{a}{}{,} 
\newcommand{\lawler}{Lawler \& Sneden}

\begin{document}
   \title{Partition functions and equilibrium constants for diatomic molecules and atoms of astrophysical interest \thanks{Full tables 1, 2, 4, 5, 6, 7 and 8 are only available in electronic form at the CDS via anonymous ftp to cdsarc.u-strasbg.fr (130.79.128.5) or via http://cdsweb.u-strasbg.fr/cgi-bin/qcat?J/A+A/.}}
   \titlerunning{Partition functions for molecules and atoms}

   \author{P. S. Barklem\inst{1}
          \and
          R. Collet\inst{2,3}
          }

   \institute{Theoretical Astrophysics, Department of Physics and Astronomy, Uppsala University,
              Box 516, SE-751 20 Uppsala, Sweden  
         \and
              Research School of Astronomy and Astrophysics, Australian National University, Canberra, ACT 2611, Australia
          \and Stellar Astrophysics Centre, Department of Physics and Astronomy, Aarhus University, Ny Munkegade 120, DK-8000 Aarhus C, Denmark
             }

   \date{Received 14 July 2015 ; accepted 28 January 2016}

 
  \abstract{Partition functions and dissociation equilibrium constants are presented for 291 diatomic molecules for temperatures in the range from near absolute zero to $10\,000$~K, thus providing data for many diatomic molecules of astrophysical interest at low temperature.  The calculations are based on molecular spectroscopic data from the book of Huber and Herzberg with significant improvements from the literature, especially updated data for ground states of many of the most important molecules by Irikura.  Dissociation energies are collated from compilations of experimental and theoretical values.  Partition functions for 284 species of atoms for all elements from H to U are also presented based on data collected at NIST.  The calculated data are expected to be useful for modelling a range of low density astrophysical environments, especially star-forming regions, protoplanetary disks, the interstellar medium, and planetary and cool stellar atmospheres.  The input data, which will be made available electronically, also provides a possible foundation for future improvement by the community.  }

   \keywords{molecular data --- atomic data}

   \maketitle
%

\section{Introduction}

Calculations of the relative abundances of molecules and atoms are required for the modelling of a wide range of cool astronomical objects and their spectra, e.g.\ the solar atmosphere, sunspots, cool stellar atmospheres, exoplanetary atmospheres and comets; for a recent review see \cite{2009IRPC...28..681B}.  In 1979, Huber and Herzberg published their classic compilation of constants for diatomic molecules \citep[][hereafter HH]{HH}.  Soon after, \citet[][hereafter ST]{1984ApJS...56..193S} published a set of partition functions and equilibrium constants for 301 diatomic molecules of possible astrophysical interest based on data from this compilation.  They calculated for temperatures in the range from $1\,000$ to $9\,000$~K, arguing that molecular equilibrium will not be valid in astrophysical objects at lower temperatures.  Nevertheless, molecular equilibrium is still often useful, either to use directly in exploratory models, or as an initial guess for non-equilibrium modelling.  For example, \citet{2010ApJ...715.1050B} have modelled terrestrial planet formation assuming chemical equilibrium, supported by arguments regarding the relevant timescales for planetesimal formation and observational evidence from our solar system.  In this work we redo the calculations of ST, with the main goal to extend the temperature range of the calculations down to temperatures near absolute zero through removal of various high-temperature approximations related to the rotational and nuclear hyperfine structure levels.   

These calculations allow us to make some additional improvements.  First, since the HH data is now made available electronically via the NIST\footnote{National Institute of Standards and Technology, \url{http://www.nist.gov/}} Chemistry Webbook \citep{HHb}, all states and coefficients could be included with little effort.  Second, critically compiling input molecular data is a complex problem, one an expert in molecular physics could spend a career on: Huber and Herzberg spent over 30 years making their compilation.  Thus, ST chose to adopt data exclusively from HH so that there could be no ambiguity regarding what data was used, even though this meant that improvements from after HH finalised data for a given molecule were omitted.  This problem can now be relatively easily circumvented by making the input data files available electronically, allowing data sets to be merged in a transparent fashion.   Finally, the possibility to provide large tables online means results can be presented for a fine temperature grid and removes the need for polynomial representations.

There have been countless improvements and additions to known molecular data during the over 30 years since HH's compilation.  The dissociation energy is of particular importance in the calculation of the equilibrium constant, and so in this work we have compiled updated dissociation energies based on two main sources: the compilation of experimental bond energies of \cite{Luo}, and the theoretical calculations of \cite{1991JChPh..94.7221C} using Gaussian-2 theory.  Regarding spectroscopic constants, the most significant update was the inclusion of constants for the ground states of 85 of the most important molecules according to the compilation by \cite{2007JPCRD..36..389I}; this addition is of particular importance in updating the HH data for this work.  Further, in the case of a few molecules of particular astrophysical interest for which data did not previously exist (e.g. FeH, TiH) or improved data has now become available (e.g. CrH, ZrO), data were added or updated, respectively.  We emphasise that we have not attempted here to make a comprehensive review of all 291 molecules; however, the input data presented here provides a possible foundation for workers to be able to incorporate further improvements for specific molecules of particular interest.  Our primary goal here is to provide a comprehensive data set for a large number of molecules suitable for astrophysical modelling, using up-to-date, or at least reasonably close to up-to-date, data in the most important cases. A few molecules were also removed from the list of ST, and the final list contains 291 diatomic molecules; all improvements and removals with respect to ST, are detailed in \S~\ref{sect:specconsts} .

In order to calculate equilibrium constants, in addition to partition functions for the molecule, partition functions for the component atoms are needed.  These partition functions are of considerable interest in themselves, and have also been calculated based on data from NIST for all chemical elements from hydrogen to uranium, for neutral and singly- and doubly-ionised species. 

The paper is structured as follows.  In Section~\ref{sect:calc} the calculations and the input data are described in detail.  In Section~\ref{sect:results} the results are analysed and compared with previous results.

\section{Calculations}
\label{sect:calc}

The expressions used for calculating the partition function corresponds to those from \cite[eqns. 26--31]{1987A&A...182..348I}, though neglecting spin-orbit coupling. We write the partition function as:
\begin{eqnarray}
Q & = & 
\sum_e \sum_{\upsilon=0}^{\upsilon_\mathrm{max}} \sum_{J=\Lambda}^{J_\mathrm{max}} g_{\Lambda, \mathrm{hfs}} (2S+1) (2J+1) \nonumber \\
&& \times \exp \left[ - \frac{hc}{kT} \left( T_e + G(\upsilon) + F_\upsilon(J) - E_0 \right) \right],
\label{eqn:1}
\end{eqnarray}
where $e$ denotes electronic states with projection of orbital angular momentum quantum number $\Lambda$ and total spin quantum number $S$, and $\upsilon$ and $J$ are the vibrational and rotational quantum numbers, respectively.  $T_e$, $G$ and $F$ represent the electronic, vibrational and rotational energies in wavenumber units.  $E_0$ is the energy of the lowest state of the molecule (i.e. $\upsilon = 0$, $J=\Lambda$ for the ground electronic state); we note that although this term can be taken outside the summation, at low temperatures it is best left in the summation to avoid numerical errors.  



The vibrational and rotational energies can be expressed as asymptotic expansions in terms of molecular constants.  In this work we use the usual band spectrum constants as defined in \cite{1950msms.book.....H} and used by HH.  The vibrational energy is
\begin{equation}
G(\upsilon) = w_e(\upsilon+\frac{1}{2}) - w_e x_e (\upsilon+\frac{1}{2})^2 + w_e y_e (\upsilon +\frac{1}{2})^3.
\label{eqn:2}
\end{equation}
The rotational energy is
\begin{equation}
F_\upsilon(N) = B_\upsilon J(J+1) - D_\upsilon (J(J+1))^2,
\label{eqn:3}
\end{equation}
where
\begin{equation}
B_\upsilon = B_e - \alpha_e (\upsilon+\frac{1}{2}) + \gamma_e (\upsilon+\frac{1}{2})^2,\end{equation}
with the centrifugal distortion term
\begin{equation}
D_\upsilon = D_e - \beta_e (\upsilon+\frac{1}{2}).
\end{equation}
The statistical weight factor relating to $\Lambda$-doubling and nuclear hyperfine structure $g_{\Lambda, \mathrm{hfs}}$ is calculated according to \citet[][Table~3 and eqns.~28 and~29]{1987A&A...182..348I}.  Note that our calculations follow the usual convention that the statistical weight is divided by the product of the nuclear spin statistical weights, i.e. nuclear spin parts are not included.

The upper limit $\upsilon_\mathrm{max}$ for summation over vibrational quantum number $\upsilon$ in eqn.~\ref{eqn:1}, is found naturally by treating eqn.~\ref{eqn:2} as an asymptotic expansion.  Normally, the best approximation is achieved when the asymptotic expansions are truncated at the smallest term.   Due to the small number of terms available, to ensure the convergence of the included terms we used a slightly more stringent condition: to be included in the summation a term should not be more than half of the preceding term.  In cases where $w_e x_e$ is not available or able to be estimated, the summation is arbitrarily truncated at $\upsilon_\mathrm{max}=100$.  For the summation over rotational quantum number $J$, the value of $J$ with the largest thermal population can be easily calculated \citep[see pg. 124 of][]{1950msms.book.....H}.  As the population decreases exponentially at larger $J$, we chose to truncate the summation at $J_\mathrm{max}$ equal to 10 times this value, but with a minimum value of 10, which is used at low temperatures when $J=0$ is the most populated state (for $\Sigma$ states).

The main differences between our calculations and those of ST are: a) we have not employed the high-temperature approximation for the summation over rotational quantum number $J$, b) we have not used the approximate expression for the statistical weight $g$, c) we include higher order terms in the asymptotic expansions if available.  Both a) and b) are necessary to extend the calculations to low temperatures, as both these approximations are only valid in the high-temperature case where many rotational and nuclear hyperfine levels contribute to the partition function.

As mentioned earlier, in order to compute the molecular equilibrium constants, we need to evaluate partition functions for the relevant atoms and ions as well.  We calculate them using the following expression:
\begin{equation}
Q_\mathrm{atom} = \sum_i\,\left(2J_i+1\right)\,e^{-\chi_i/kT}
\label{eq:qatom}
\end{equation}
where the index $i$ refers to the atomic energy level, $J_i$ is the angular momentum of the level and $\chi_i$ its excitation energy relative to the ground state. 
In the calculations, we account explicitly for fine and hyperfine structure of atomic levels and again follow the usual convention that the nuclear spin statistical weights are not included.  

For isolated atoms and molecules the partition function formally diverges due to an infinite number of Rydberg states approaching the ionisation threshold.  However, in real physical environments the sum is modified due interactions of nearby particles.  This may be viewed in the so-called ``physical'' or ``chemical'' pictures \citep[e.g.][]{Dappen1987,rogers_opal_1996}.  In the physical picture, nearby particles modify the electronic potential from the spherically symmetric Coulomb potential, thus resulting in a finite number of states.   In the chemical picture, one begins with the isolated atoms and molecules and account for loosely bound states being ``dissolved'' into continuum states if a perturbing particle is sufficiently nearby.   Inclusion of these effects would lead to partition functions and equilibrium constants dependent on densities of the most common perturbers (hydrogen atoms, ions, electrons, etc.), leading to significantly increased complexity, and perhaps limiting the usefulness, of the results.   In these calculations we have included all available states from the relevant sources, without modification, and will show in Sect.~\ref{sect:results} that this leads to only small errors for typical low density astrophysical environments where this data might be applied, such as stellar atmospheres.

Once partition functions for the molecule and constituent atoms and ions are known, the equilibrium constant $K_\mathrm{AB}$ can be easily calculated. In terms of partial pressures \citep[e.g.][]{1966PDAO...13....1T}  
\begin{equation}
^p K_\mathrm{AB} = \frac{p_\mathrm{A} p_\mathrm{B}}{p_\mathrm{AB}} = \left( \frac{2\pi mkT}{h^2} \right)^{3/2} kT \frac{Q_\mathrm{A} Q_\mathrm{B}}{Q_\mathrm{AB}} e^{-D_0^0/kT} .
\label{eqn:equil}
\end{equation}
The partition function and equilibrium constant calculations were carried out as detailed above using a computer program written in IDL.

\subsection{Input molecular data}

The input data for the molecular partition function calculations consists of molecular terms containing relevant quantum numbers $\Lambda$ and $S$ and spectroscopic constants $T_e, w_e, w_e x_e, w_e y_e, B_e, \alpha_e, \gamma_e, D_e, \beta_e$.  In the homonuclear case, parity (u or g) and reflection symmetry ($+$ or $-$), as well as the nuclear spins for the component atoms, $I_1$ and $I_2$, are also needed in the calculation of $g_{\Lambda, \mathrm{hfs}}$.  Calculation of the equilibrium constant requires the dissociation energy $D_0^0$.  Below we detail the data used in our calculations.

\subsubsection{Dissociation energies}

The exponential term in equation~\ref{eqn:equil} containing the dissociation energy $D_0^0$ arises since the atomic and molecular partition functions are calculated with reference to different zero points, their respective ground states.  The exponential dependence on $D_0^0$ means $K_\mathrm{AB}$, and thus the partial pressure $p_\mathrm{AB}$, number density $n_\mathrm{AB}$ and opacity $\kappa_\mathrm{AB}$, in an astrophysical model, are sensitive to errors in $D_0^0$.  The relative error from this source in the partial pressure or number density of the molecule, and thus its opacity $\kappa_\mathrm{AB}$, is given by
\begin{equation}
\frac{\Delta n_\mathrm{AB}}{n_\mathrm{AB}} = \frac{\Delta p_\mathrm{AB}}{p_\mathrm{AB}} = \frac{\Delta \kappa_\mathrm{AB}}{\kappa_\mathrm{AB}} = \exp(\Delta D_0^0/kT) - 1. 
\end{equation}
For example, at 3500~K $kT\sim 0.3$ eV and so $\Delta D_0^0 \sim 0.1$~eV leads to errors of around 40\%.  The error increases as temperature decreases; for example at 1000~K, the same error in $D_0^0$ leads to an error in excess of a factor of 2.

Thus, it is important to have accurate dissociation energies, and we made a significant effort to collect updated data.   Data were collected from three sources: HH, \cite{Luo}, and  \cite{1991JChPh..94.7221C}.  The data from the compilation of HH is compiled as a starting reference point, but is rarely adopted (only for HeH$^+$, NeH$^+$, HF$^+$). \cite{Luo} presents a comprehensive survey of experimental chemical bond energies, from which $D_0^0$ can be easily calculated.  For each molecule, \cite{Luo} presents a survey of available data, and often recommends one of the values.  If a recommended or only a single value is available, then this is was the value collected.  If no value was recommended we collected the value with the smallest estimated uncertainty.  Finally, data from the theoretical calculations of \cite{1991JChPh..94.7221C} were also collected.  This work employs so-called Gaussian-2 theory, essentially \emph{ab initio} molecular structure calculations using Gaussian orbitals; see \cite{1991JChPh..94.7221C}.

We then evaluated the data in each case and adopted a final value.  In the vast majority of cases, the value from \cite{Luo} was adopted.  The \cite{Luo} survey usually includes the experimental data in HH together with error estimates.  As error estimates were rarely given by HH, \cite{Luo} is always preferred over HH even if the data are from the same original source.  In the majority of cases, the \cite{Luo} values and the \cite{1991JChPh..94.7221C} values agreed well, and where this was the case we favoured the \cite{Luo} values as they are experimental and have error estimates.  However, in cases where disagreement was substantial, and the available experimental data in \cite{Luo} were deemed uncertain, we have favoured the theoretical values from \cite{1991JChPh..94.7221C} \citep[see discussion in][]{1994LNP...428..250C}.  

\begin{figure*}
   \centering
   \includegraphics[width=180mm,angle=180]{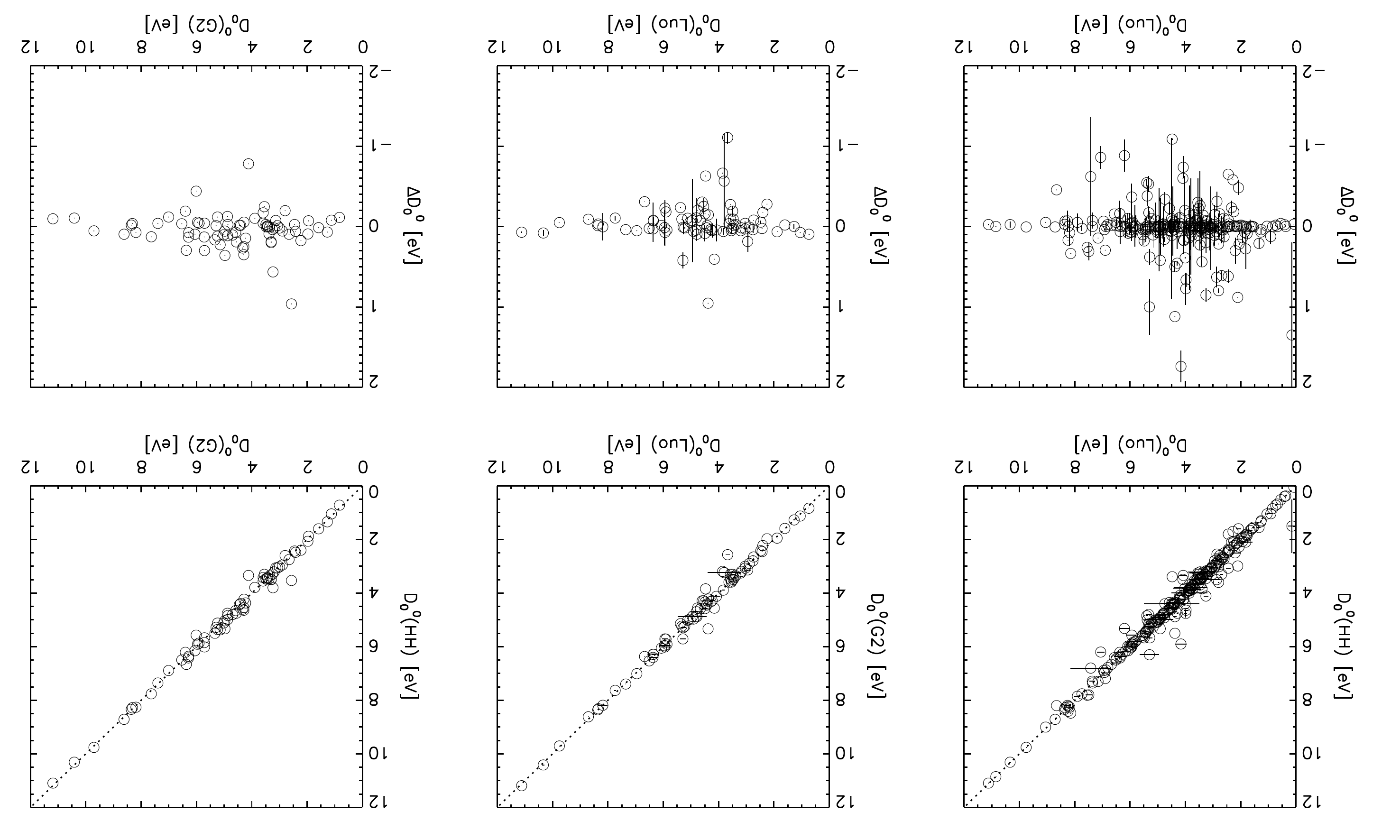}
      \caption{Comparison of dissociation energies $D_0^0$ from the three main sources discussed in the text. The bars show estimated errors where available.  The theoretical values of \cite{1991JChPh..94.7221C} are labelled G2, denoting Gaussian-2 theory. The difference shown in the lower panel is always in the sense (y-axis $-$ x-axis).  The error shown in the difference is taken as the sum of the errors, where available.
              }
         \label{fig:diss}
   \end{figure*}
\begin{figure}
   \centering
   \includegraphics[width=90mm,angle=0]{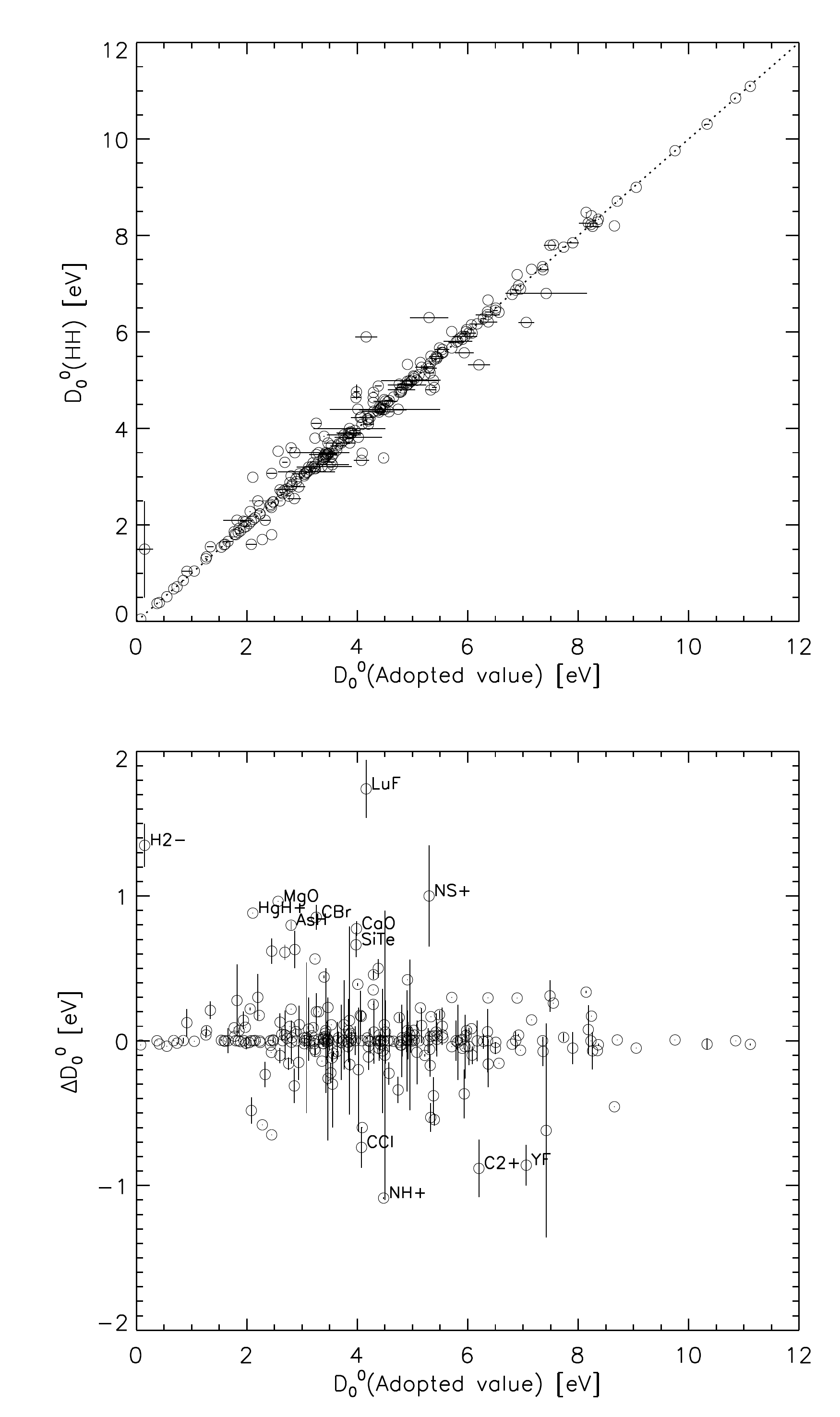}
      \caption{Comparison of final adopted values of the dissociation energies $D_0^0$ with those of HH.  The bars show estimated errors where available.  The difference is in the sense (HH $-$ adopted).  The error shown in the difference is taken as error in the adopted value, where available.  The cases with the largest differences are labelled.
              }
         \label{fig:diss2}
   \end{figure}

Data from the three different sources and the adopted data are presented in Table~\ref{tab:diss}.  Comparisons of the data, where they overlap, are given in Fig.\ref{fig:diss}.  We see that the correlation is generally good between the three sources, yet the scatter is generally of order a few 0.1 eV, and some cases disagreeing by 1 or even 2 eV.  Fig.~\ref{fig:diss2} compares the final adopted values with those from HH.   We see rather large changes in molecules such as H$_2^-$ (which we note is still rather uncertain), MgO \citep[which was noted by][]{1994LNP...428..250C}, CaO, NH$^+$, NS$^+$ and C$_2^+$, just to pick out some of the largest with constituent atoms that are astrophysically abundant.  

\begin{table*}
\begin{center}
\caption{Sample of electronic table containing dissociation energies from the literature, and final adopted values. The full table is available electronically at CDS.}
\label{tab:diss}
\scriptsize
\begin{verbatim}
Dissociation energies         291 molecules
       Molecule                    HH                   Luo                    G2               Adopted
                        Do  sigma(Do)         Do  sigma(Do)         Do  sigma(Do)         Do  sigma(Do)
                      [eV]       [eV]       [eV]       [eV]       [eV]       [eV]       [eV]       [eV]
   H2    H    H    4.478130         .    4.478007  0.000004           .         .    4.478007  0.000004
  Li2   Li   Li    1.046000         .    1.049900         .    1.124000         .    1.049900         .
   B2    B    B    3.020000         .    2.802000         .           .         .    2.802000         .
   C2    C    C    6.210000         .    6.371000  0.160000    6.401000         .    6.371000  0.160000
   N2    N    N    9.759400         .    9.753940  0.000900    9.705000         .    9.753940  0.000900
   O2    O    O    5.115600         .    5.116420  0.000130    5.013000         .    5.116420  0.000130
   F2    F    F    1.602000         .    1.605960  0.001000    1.587000         .    1.605960  0.001000
...
\end{verbatim}
\end{center}
\end{table*} 

\subsubsection{Spectroscopic constants}
\label{sect:specconsts}

The starting point for our calculations are the molecular constant data from HH, which were extracted electronically from the NIST Chemistry WebBook \cite[][retrieved October 14, 2011]{HHb} for all molecules considered by ST.  An IDL computer program was then used to automatically extract the required data from these html files.   Changes and additions to the data are then made by merging the data with extra data sets, also with an IDL computer program.  The most important of these changes and additions is the compilation of updated constants for the ground states of 85 molecules by \cite{2007JPCRD..36..389I, 2009JPCRD..38..749I}.  Further changes and additions were also made as listed below.  Note, for negative ions we followed the treatment of ST (see \S~IIa of that paper) adopting estimates of data and ground state designations from similar molecules where necessary.

\paragraph{Additional molecules}    TiH and FeH were added using data from \cite{2005ApJ...624..988B} and \cite{2003ApJ...594..651D}, respectively.  CH$^-$ was also added, which is not present in Table 6 of ST despite being mentioned in \S~IIa of that paper.  In this case we took the molecular term for the ground state from HH, and used spectroscopic constants from the CH ground state.

\paragraph{Updates to data}  The constants for CrH have been updated using data from \citet[][Table II]{2001JChPh.115.1312B}, and from \cite{2004JChPh.121.8194G} for higher states, prefering CASPT2 calculations.  The energy levels for ZrO have been updated following \cite{1988ApJ...332.1090D}.  The constants for the ground state of NeH$^+$ have been taken from \cite{civis_new_2004}.

\paragraph{Estimates}  Data are missing for some states with identified quantum numbers and electronic energies $T_e$, and so rather than omit these states, in certain cases we chose to make simple estimates based on data from other states of the molecule or similar molecules.  In cases where $w_e$ or $B_e$ is missing, but available for other states of the molecule, we adopted the average of the values for the other states.  In cases where  $w_e$ or $B_e$ is missing for the ground state of a molecule that has no other states with these values, then an estimate was taken from a similar molecule (e.g. oxide or homo-nuclear molecule with nearest mass) where data is available for the ground state.  

\paragraph{Removed molecules}  Some molecules included in ST, were not included in this work.  PdH, ClF, AuSi, CuSe were missing in the NIST Webbook and not deemed important enough to warrant followup.  HeH$^+$, CH$^+$ and LaO were also missing, but included using data from HH added by hand.  LaH, BH$^+$, RuO, AgO, CeO, IrO, FeF, MnS, FeCl either do not have an identified ground state or do not have data for the ground state and were not deemed important enough to warrant followup.  It seems that ST assumed a ground state for these cases. For CN$^+$ the ground state is also not identified in HH, but due to its possible astrophysical importance we chose in this case to assume that the lowest known state, $a ^1 \Sigma$, was the ground state.  We note the inherent danger in such a procedure:  e.g., the ground state of FeCl is uncertain in HH, and later work by \cite{0022-3700-13-24-010} showed it to be a state with different properties to those identified as candidates in HH.

The final input data for the calculations are written to files containing all relevant information, including dissociation energies, nuclear spins, and spectroscopic constants.  Labels indicating the source of the data are included for all spectroscopic constants.  These files are available in electronic form from the CDS.  A sample of an example file for H$_2$ is given in Table~\ref{tab:input}.  The key to the reference labels used in these files is given in Table~\ref{tab:inputrefs}.  We note that states where fine structure components are given in HH, with corresponding projection of total angular momentum quantum number $\Omega$, the value of the spin multiplicity $2S+1$ is adjusted to unity in order to achieve the correct statistical weight.

\begin{table*}
\begin{center}
\caption{Sample of electronic table for H$_2$ containing adopted dissociation energy, nuclear spins, and spectroscopic constants. The full table is available electronically at CDS.}
\label{tab:input}
\scriptsize
\begin{verbatim} 
       D0        I1        I2
 4.4780070       0.5       0.5
                      label   name Lambda   2S+1    +/-    u/g           Te           we          wxe    ...   references 
                                                                     [cm-1]       [cm-1]       [cm-1]    ...                 
...
           C ^1 Pi_u  2p-pi      C      1      1      .      u     100089.8      2443.77       69.524    ...   HH     HH     HH     ...
  B ^1 Sigma_u ^+  2p-sigma      B      0      1      +      u      91700.0      1358.09       20.888    ...   HH     HH     HH     ...
X ^1 Sigma_g ^+  1s-sigma^2      X      0      1      +      g            0    4401.2130    121.33600    ...   HH     IR     IR     ...
END
\end{verbatim}
\end{center}
\end{table*} 

\begin{table}
\begin{center}
\caption{Key to reference labels in electronic files.  The full details of each are given in the reference list of this paper.}
\label{tab:inputrefs}
\scriptsize
\begin{verbatim} 
B01  = Bauschlicher et al (2001)
B05  = Burrows et al (2005)
C04  = Civis et al (2004)
D03  = Dulick et al (2003)
DH88 = Davis and Hammer (1988)
DS82 = Delaval and Schamps (1982)
EST  = Estimated from other states or from similar molecule
G04  = Ghigo, Roos, Stancil and Weck (2004) 
HH   = Huber and Herzberg (1979)
IR   = Irikura (2007)
\end{verbatim}
\end{center}
\end{table} 

\begin{table}
\begin{center}
\caption{Sample of electronic table with first, second, and third ionisation energies for atoms and ions (for H and He, the highest possible ionisation stages are singly and doubly ionised, respectively, and therefore the ionisation energy of higher stages are meaningless and are assigned a value of $-1$ in the table). The full table is available electronically at CDS.}
\label{tab:ionpot}
\scriptsize
\begin{verbatim} 
* Atom.Num.  Elem.     IE1          IE2         IE3 
*                      [eV]         [eV]        [eV]
      1        H      13.5984      -1.000      -1.000
      2       He      24.5874      54.418      -1.000
      3       Li       5.3917      75.640     122.454
      4       Be       9.3227      18.211     153.896
      5        B       8.2980      25.155      37.931
      6        C      11.2603      24.385      47.888
      ...
\end{verbatim}
\end{center}
\end{table} 

\begin{table}
\begin{center}
\caption{Sample of electronic table with the list of references for atomic energy level data, ion by ion. The citation keys in the table are abbreviations of the full references, which are also made available electronically in {\sc Bib}{\TeX} format. The full table is available electronically at CDS.}
\label{tab:atomrefs}
\scriptsize
\begin{verbatim}
* Elem  Ion    References (citation keys)
...
   Fe     I      Nav:1994
   Fe    II      Nav:2013
   Fe   III      Sug:1985
   Co     I      Sug:1985
   Co    II      Pic:1998  Pic:1998a   Sug:1985
   Co   III      Sug:1985
...

@article{Nav:1994,
        Author = {G. Nave and S. Johansson and R. C. M. Learner 
                   and A. P. Thorne and J. W. Brault},
        Doi = {10.1086/192079},
        Journal = {Astrophys. J., Suppl. Ser.},
        Pages = {221--459},
        Title = {A New Multiplet Table for Fe~I},
        Volume = {94},
        Year = {1994} }
...
        
\end{verbatim}
\end{center}
\end{table} 

\subsection{Input atomic data}
The input data for the atomic partition function consists of angular momenta and excitation energies of energy levels of atoms and ions for all elements between H and U and ionisation stages from I to III (neutral and singly and doubly ionised). The information was extracted electronically from the NIST atomic spectra database \citep{Kramida:2014}, which includes data about atomic levels collected and critically assembled from various compilations. The complete list of references for the atomic energy level information is provided as electronic material (a sample of which is given in Table~\ref{tab:atomrefs}). Partition functions for atoms and ions are then computed by simply combining the data according to Eq.~\ref{eq:qatom}. 

Ionisation energies for individual atoms and ions are not strictly relevant for the calculations of atomic partition functions; however, knowing their values is necessary in order to compute ionisation equilibria. For completeness and for the user's convenience, we have therefore elected to compile and make available electronically a table of ionisation energies for all elements between H and U. The ionisation energy data comes primarily from the CRC Handbook of Chemistry and Physics \citep{CRC:2010}; data missing from the primary source or affected by large uncertainties (mainly ionised stages of very heavy elements) have been filled in using theoretical  or numerically interpolated values extracted from the NIST database \citep{Kramida:2014}. A sample of the table is shown in Table~\ref{tab:ionpot}.

\section{Results and discussion}
\label{sect:results}

To present the final calculations we must first decide on a temperature grid for the final calculations.  This should ideally be as sparse as possible for presentation reasons, while retaining a sufficient number of points for precise use in applications.  To decide the temperature grid we first made a calculation on a fine logarithmic grid of 101 points ranging from 0.1 to $10^4$~K.  In implementing the results in a spectrum synthesis code, Jeff Valenti (private communication) calculated an adaptive grid that, when used with cubic spline interpolation in $\log_{10}(Q)$ for partition functions and in $\log_{10}(^pK)+(D_0\times 5040/T)$ for equilibrium constants, tabulated as a function of $\log_{10}(T)$, would provide results with relative accuracy better than $10^{-4}$ with the minimum number of points.  This grid contained 31 points.  We then constructed a temperature grid that was very close to the adaptive one, but using only round numbers, as well as adding some low temperature points to extend the grid down to $10^{-5}$~K.  This grid was then used for our final calculations, and contains 42 points.  Partition functions and equilibrium constants for molecules are presented in Tables~\ref{tab:partf} and~\ref{tab:equil}.  Partition functions for atoms are presented in Table~\ref{tab:atompartf}.  The full tables are available electronically at CDS.

\begin{table*}
\begin{center}
\caption{Sample of electronic table containing partition functions for molecules. The full table is available electronically at CDS.}
\label{tab:partf}
\scriptsize
\begin{verbatim}
Partition functions Q
         291
T [K]   1.00000e-05   1.00000e-04   ...   1.00000e+02   ...   5.00000e+03   ...   1.00000e+04

   H2   2.50000e-01   2.50000e-01   ...   6.67129e-01   ...   5.07619e+01   ...   1.94871e+02
  Li2   3.75000e-01   3.75000e-01   ...   5.26396e+01   ...   5.44859e+04   ...   2.97872e+05
   B2   1.87500e+00   1.87500e+00   ...   8.70213e+01   ...   1.86115e+04   ...   8.32502e+04
                                    ...                 ...                 ...
   CO   1.00000e+00   1.00000e+00   ...   3.64899e+01   ...   4.06593e+03   ...   1.50689e+04
                                    ...                 ...                 ...
   KI   1.00000e+00   1.00000e+00   ...   1.23175e+03   ...   1.57038e+06   ...   8.03200e+06
\end{verbatim}
\end{center}
\end{table*} 

\begin{table*}
\begin{center}
\caption{Sample of electronic table containing equilibrium constants. The full table is available electronically at CDS.}
\label{tab:equil}
\scriptsize
\begin{verbatim} 
Equilibrium constants  log10(pK)
         291
T [K]   1.00000e-05   1.00000e-04   ...   1.00000e+02   ...   5.00000e+03   ...   1.00000e+04
   
   H2  -2.25687e+09  -2.25687e+08   ...  -2.16942e+02   ...   6.59790e+00   ...   9.02320e+00
  Li2  -5.29139e+08  -5.29139e+07   ...  -4.48086e+01   ...   8.31687e+00   ...   9.39624e+00
   B2  -1.41218e+09  -1.41218e+08   ...  -1.32211e+02   ...   8.22075e+00   ...   9.77654e+00
                                    ...                 ...                 ...
   CO  -5.60286e+09  -5.60286e+08   ...  -5.50670e+02   ...   1.01177e+00   ...   6.87062e+00
                                    ...                 ...                 ...
   KI  -1.66317e+09  -1.66317e+08   ...  -1.57877e+02   ...   6.32569e+00   ...   8.72302e+00

\end{verbatim}
\end{center}
\end{table*} 

\begin{table*}
\begin{center}
\caption{Sample of electronic table containing partition functions for atoms. The full table is available electronically at CDS.}
\label{tab:atompartf}
\scriptsize
\begin{verbatim}
Partition functions Q
         284
  T [K]   1.00000e-05   1.00000e-04   ...   1.00000e+02   ...   5.00000e+03   ...   1.00000e+04
 
    H_I   2.00000e+00   2.00000e+00   ...   2.00000e+00   ...   2.00000e+00   ...   2.00015e+00
   H_II   1.00000e+00   1.00000e+00   ...   1.00000e+00   ...   1.00000e+00   ...   1.00000e+00
    D_I   2.00000e+00   2.00000e+00   ...   2.00000e+00   ...   2.00000e+00   ...   2.00014e+00
                                      ...                 ...                 ...    
   Fe_I   9.00000e+00   9.00000e+00   ...   9.01785e+00   ...   2.77940e+01   ...   5.96627e+01
                                      ...                 ...                 ...
    Cl-   1.00000e+00   1.00000e+00   ...   1.00000e+00   ...   1.00000e+00   ...   1.00000e+00
\end{verbatim}
\end{center}
\end{table*} 

\begin{figure*}
   \centering
   \resizebox{0.9\hsize}{!}{
    \includegraphics{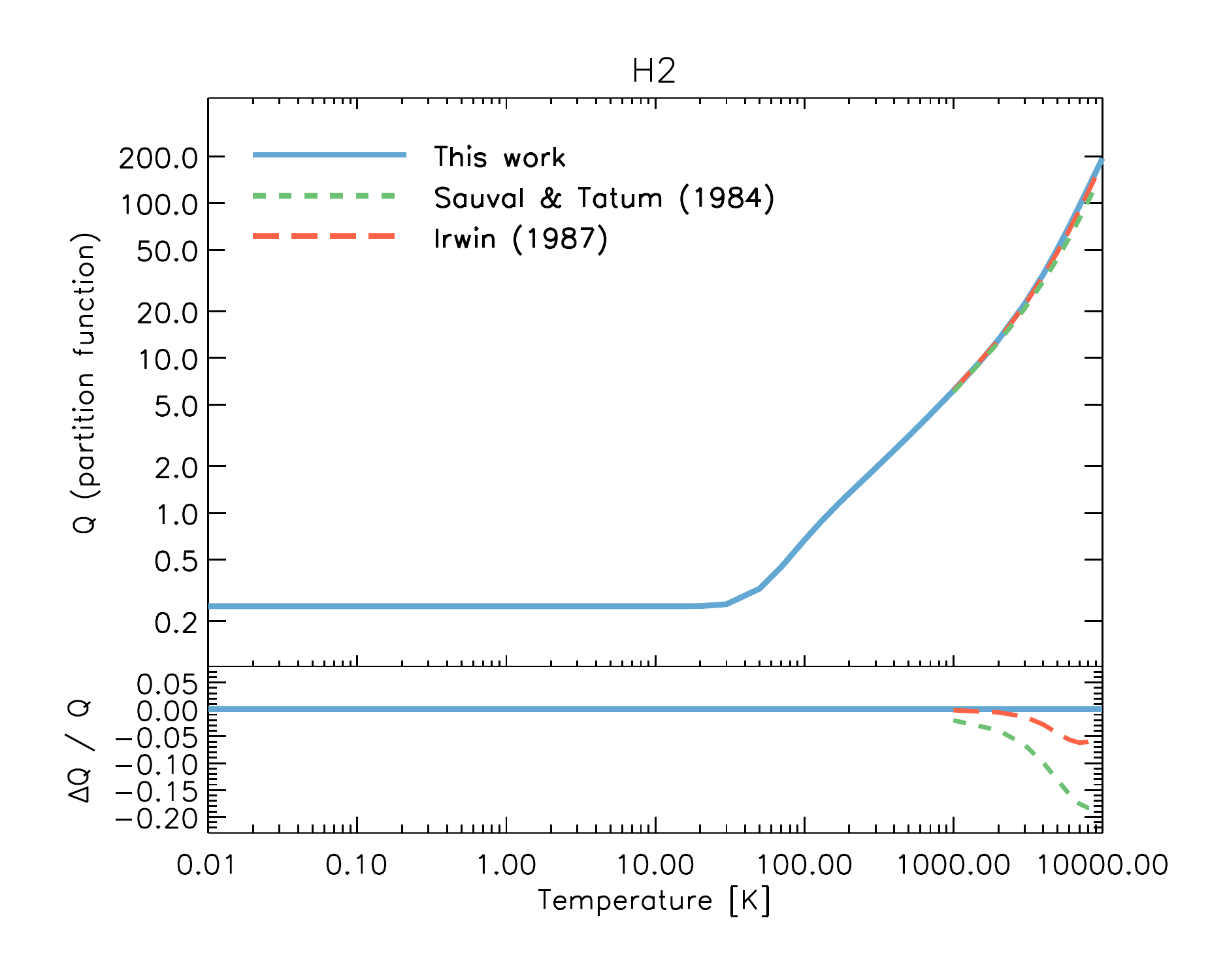} 
	\includegraphics{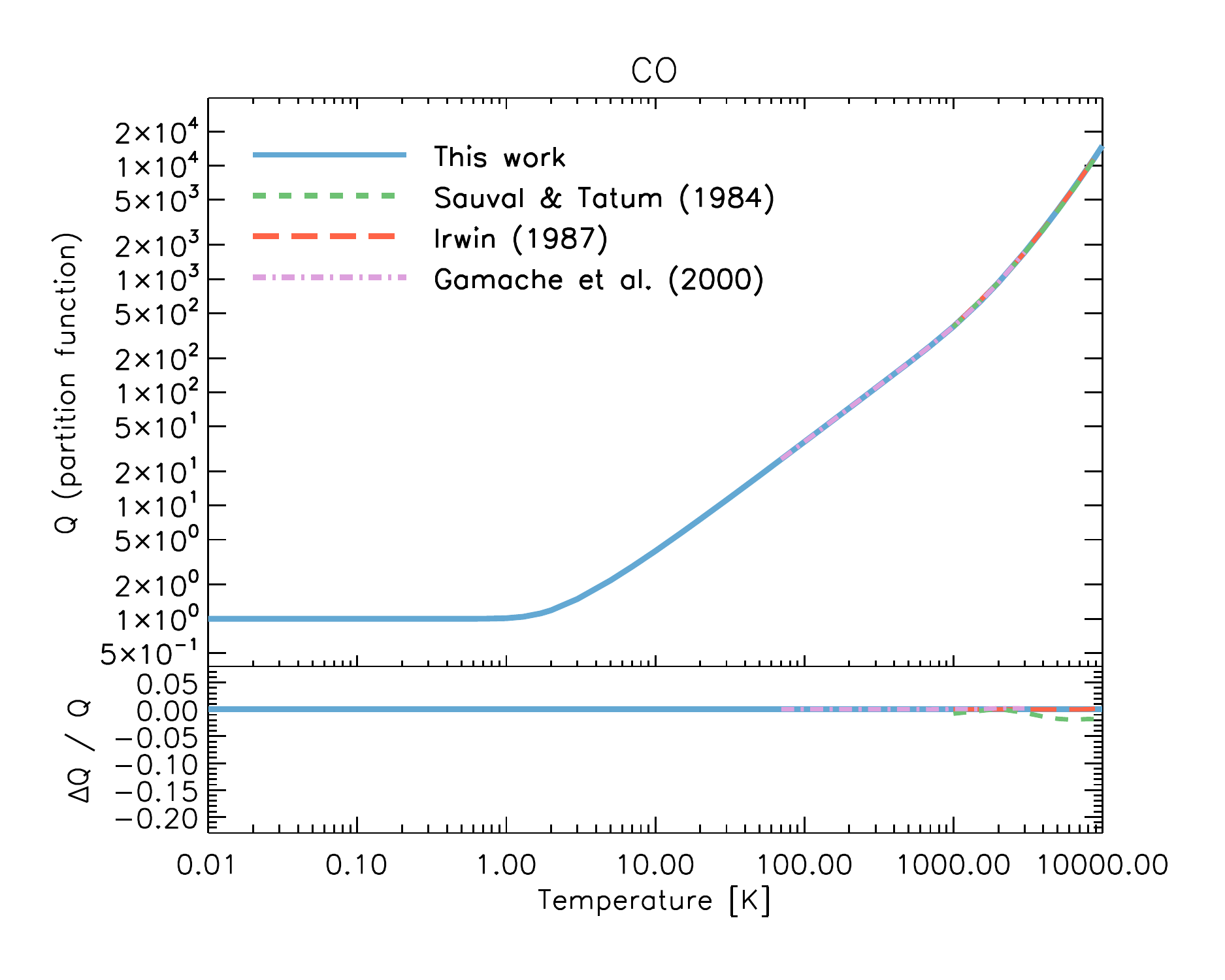} 
	}
   \resizebox{0.9\hsize}{!}{
    \includegraphics{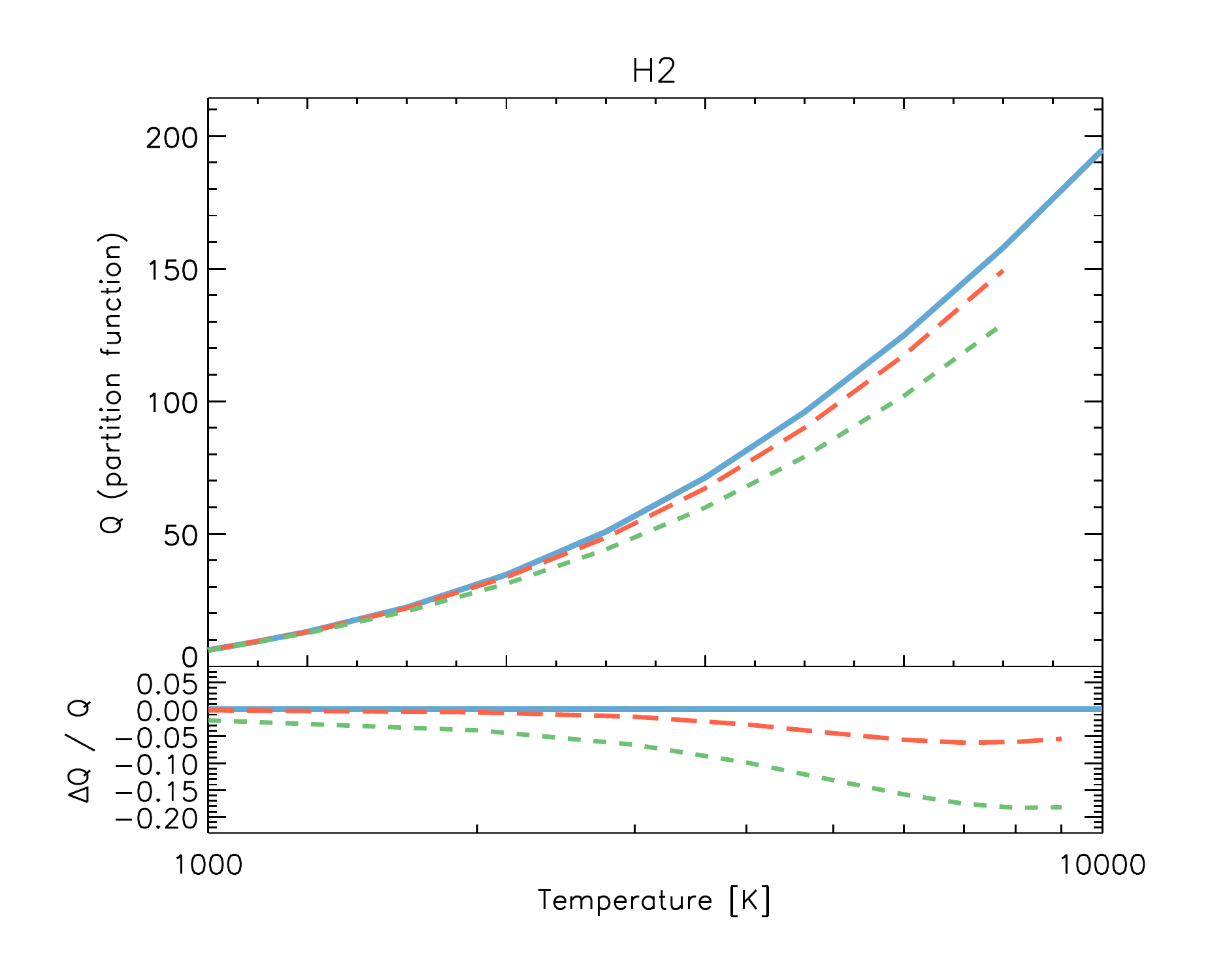} 
	\includegraphics{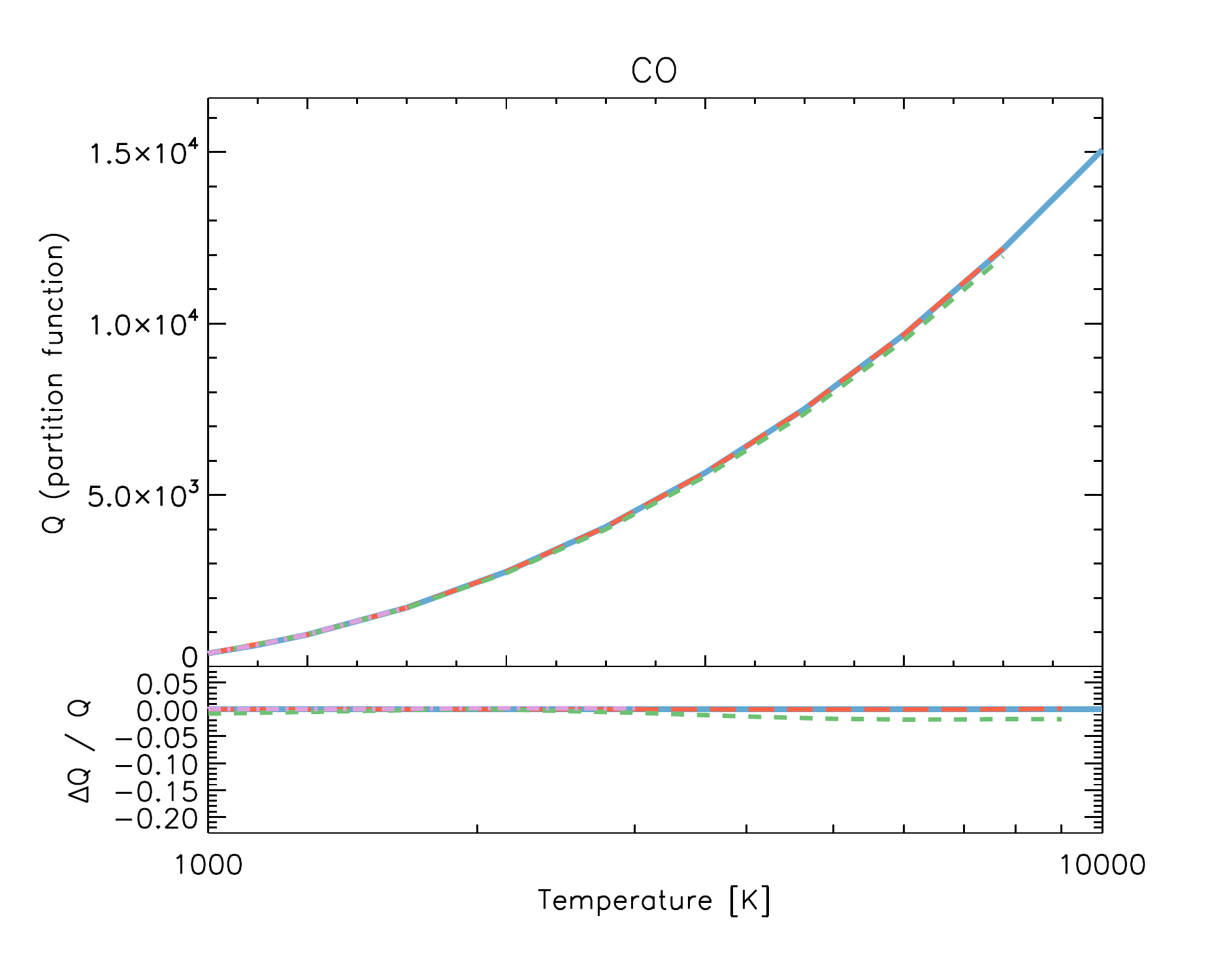} 
	}
   \resizebox{0.9\hsize}{!}{
    \includegraphics{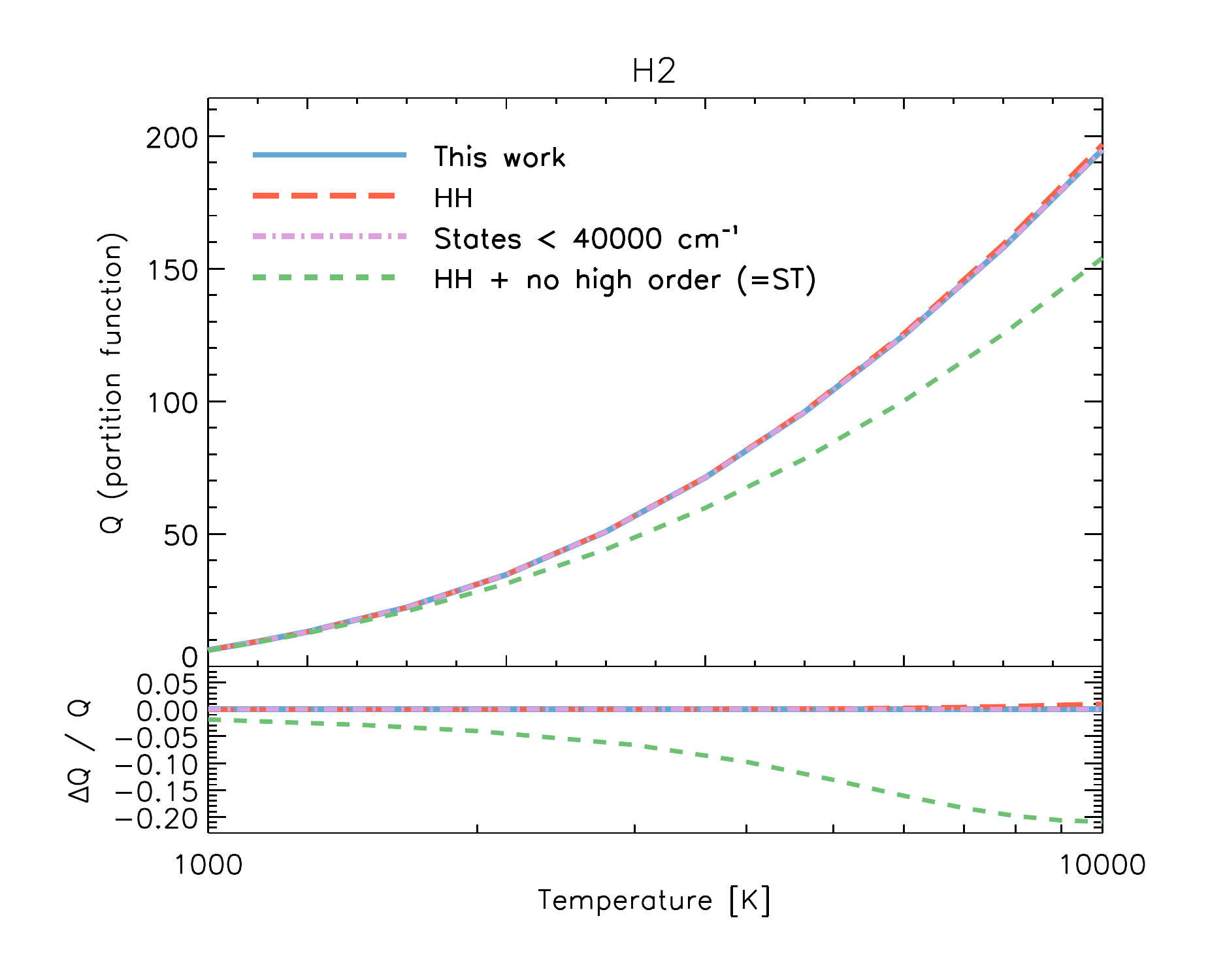} 
	\includegraphics{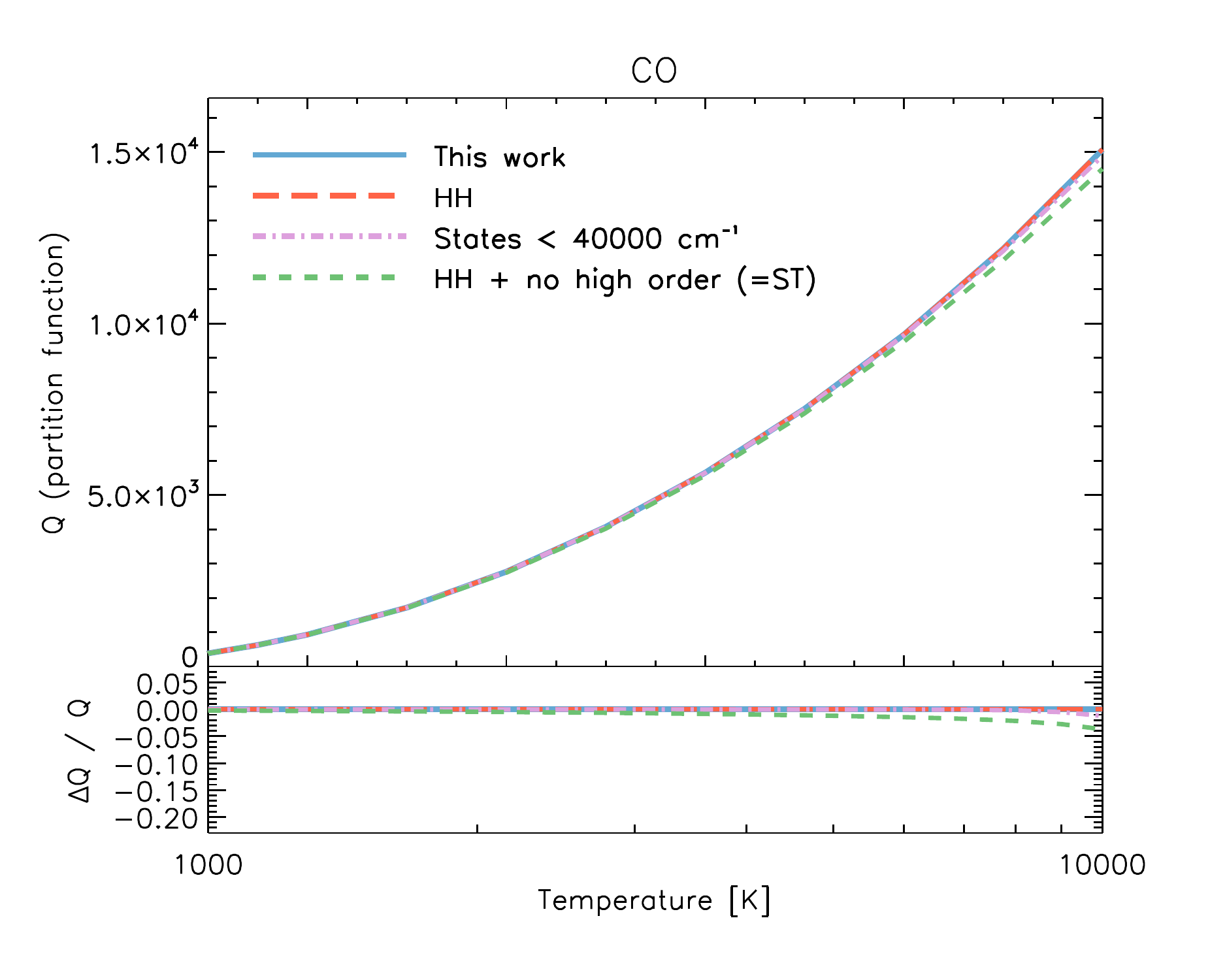} 
	}
      \caption{Partition function comparisons for  H$_2$ (\emph{left column}) and CO (\emph{right column}).  \emph{Upper row:}  The partition function ($Q$) calculated in this work (\emph{blue curves}) shown on a logarithmic scale.  Results from polynomial expressions of ST (\emph{green dashed curves}), \cite{1987A&A...182..348I} (\emph{red long-dashed curves}), and \cite{2000JMoSt.517..407G} (\emph{purple dot-dashed curves}) are also shown.  \emph{Middle row:}  The same results as for the upper panels are shown on a linear scale over the reduced temperature range $1\,000$ to $10\,000$~K.   \emph{Lower row:} Partition functions calculated in this work under various approximations, allowing the origin of any changes with respect to ST to be identified.  Partition functions are recalculated assuming constants from HH, i.e., changes and additions to molecular data are ignored (\emph{red long-dashed curves}), limiting states to those with $T_e < 40\,000$~cm$^{-1}$, as was done in ST, and thus showing the contribution of high-lying states (\emph{purple dot-dashed curves}), and with both of these approximations (neglecting high-lying states and using only data from HH), and in addition neglecting higher-order spectroscopic constants that were not considered by ST (\emph{green dashed curves}).  This final calculation should correspond to that of ST (which it does to reasonable precision by comparison with the middle panel), and allows the effect of the higher-order constants to be seen.  In each case, the \emph{lower panels} show relative differences between the plotted results and ours.
              }
         \label{fig:molpart}
   \end{figure*}

Partition functions for H$_2$ and CO, two cases of common astrophysical interest are examined in Figs.~\ref{fig:molpart} and compared with previous results from ST and \cite{1987A&A...182..348I}, and in the case of CO with \cite{2000JMoSt.517..407G}.  The main impact of the new calculations is seen in the upper and middle panels:  our calculations go down to near absolute zero temperature and there approach a constant value, specifically the statistical weight of the ground level divided by the nuclear statistical weight.  The lower panels show the impact of inclusion of high-lying states, higher-order constants and changes to the molecular data compared to ST.  We see in H$_2$ the inclusion of higher-order constants has some effect at higher temperatures, while the addition of high-lying states has no appreciable effect.  The updating of the spectroscopic constants according to \cite{2007JPCRD..36..389I} has a very small effect on the partition function.  In CO, we see none of these has a significant effect.

\begin{figure*}[htp]
\centering
\resizebox{0.9\hsize}{!}{
	\includegraphics{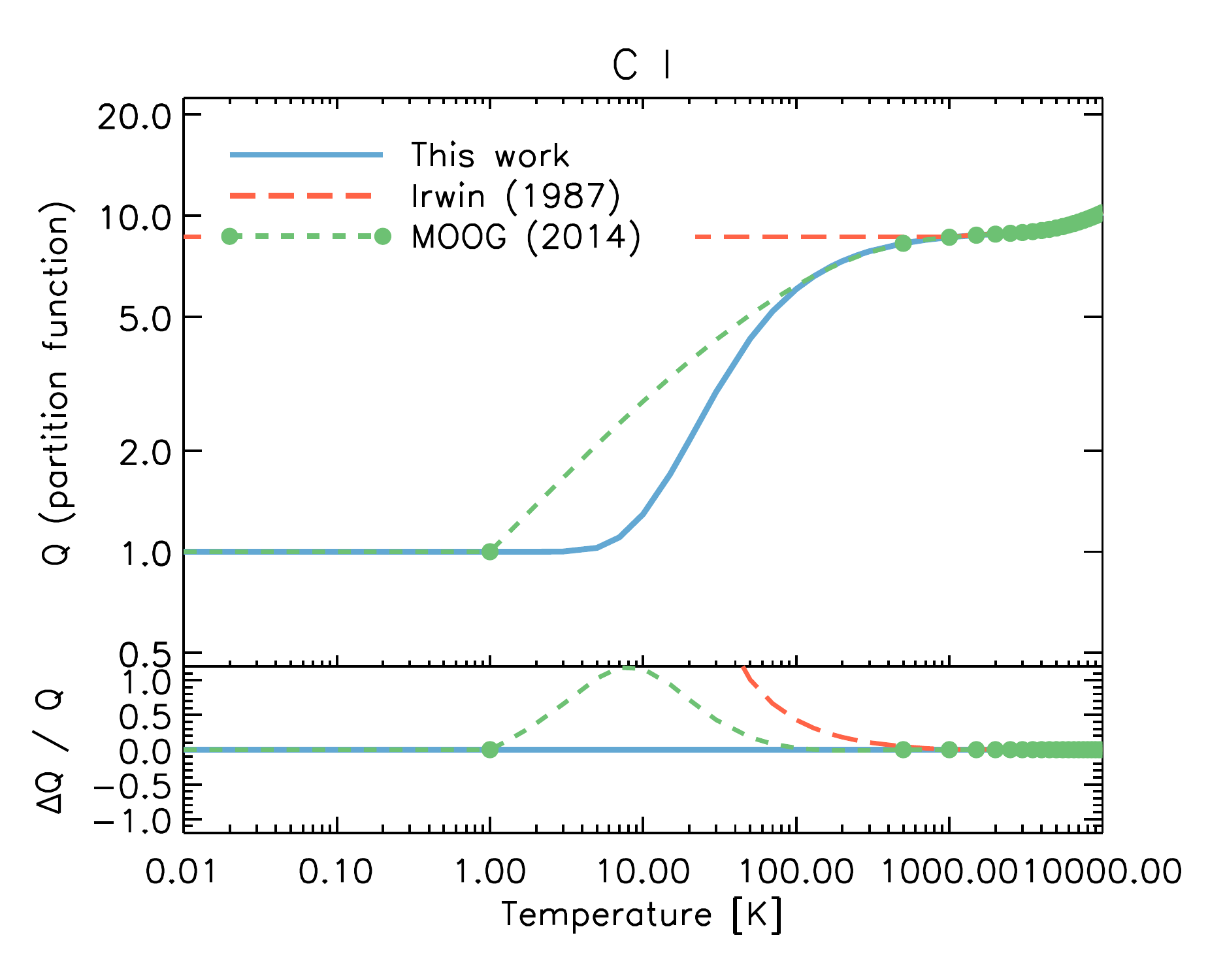}
	\includegraphics{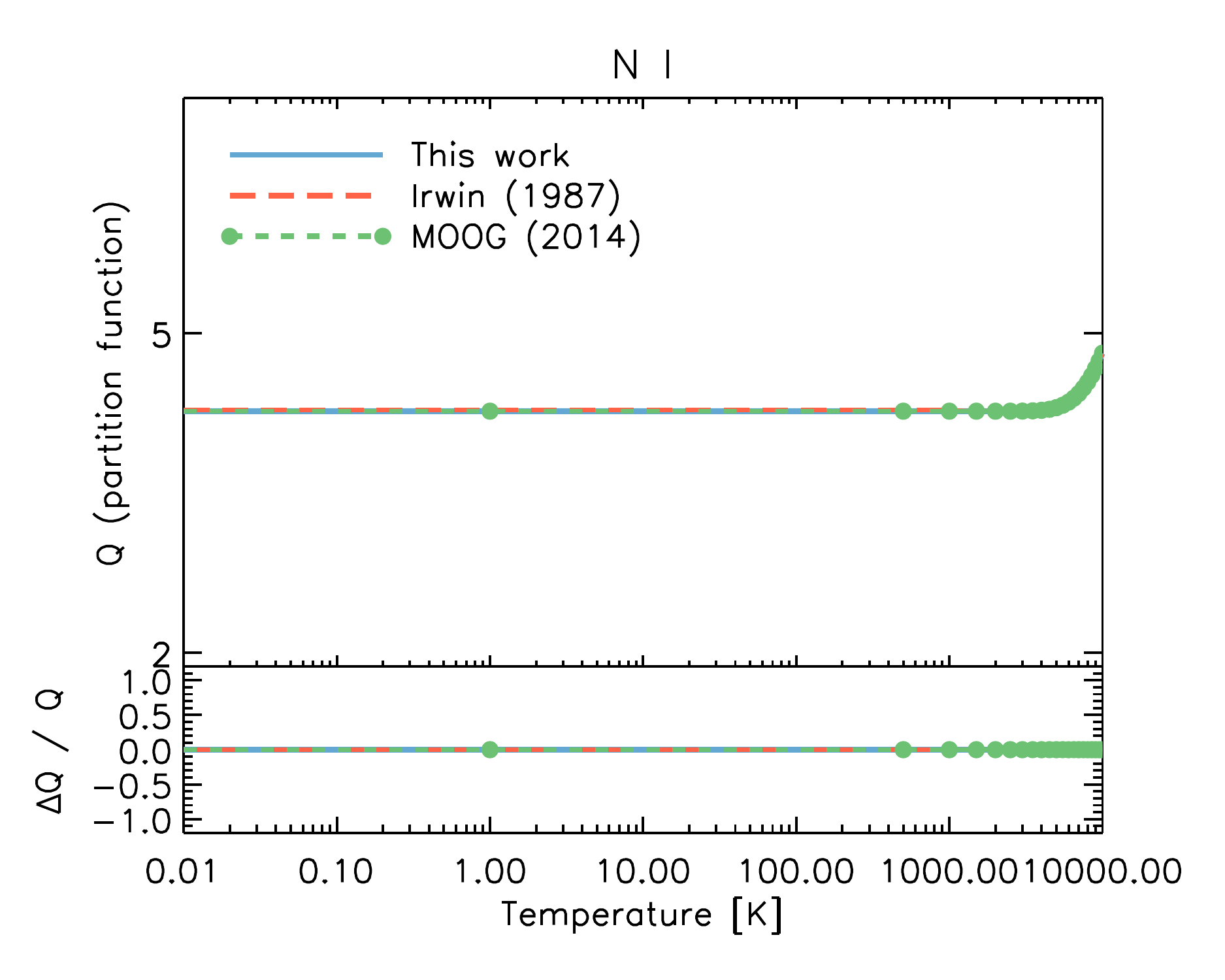}
}
\resizebox{0.9\hsize}{!}{
	\includegraphics{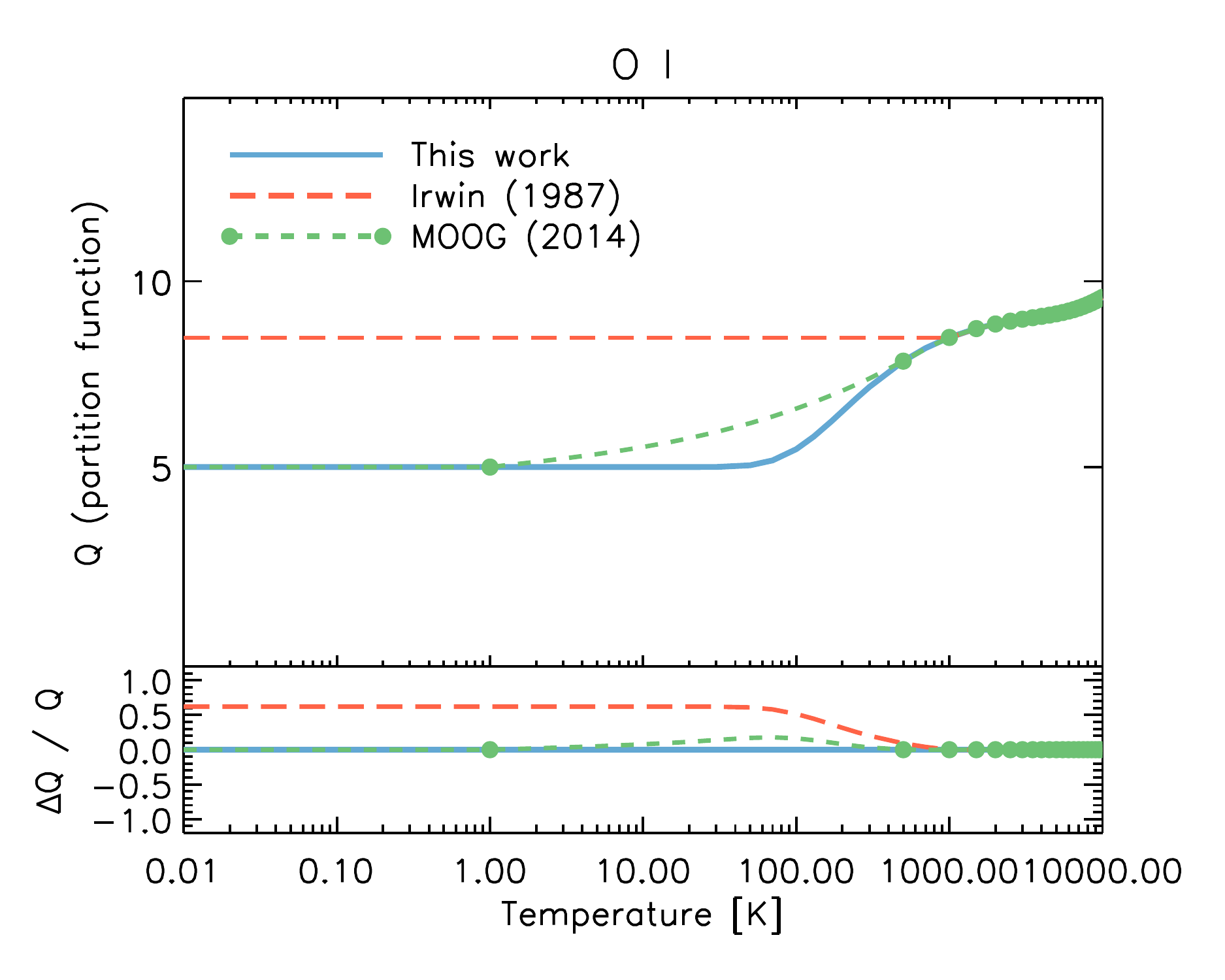}
	\includegraphics{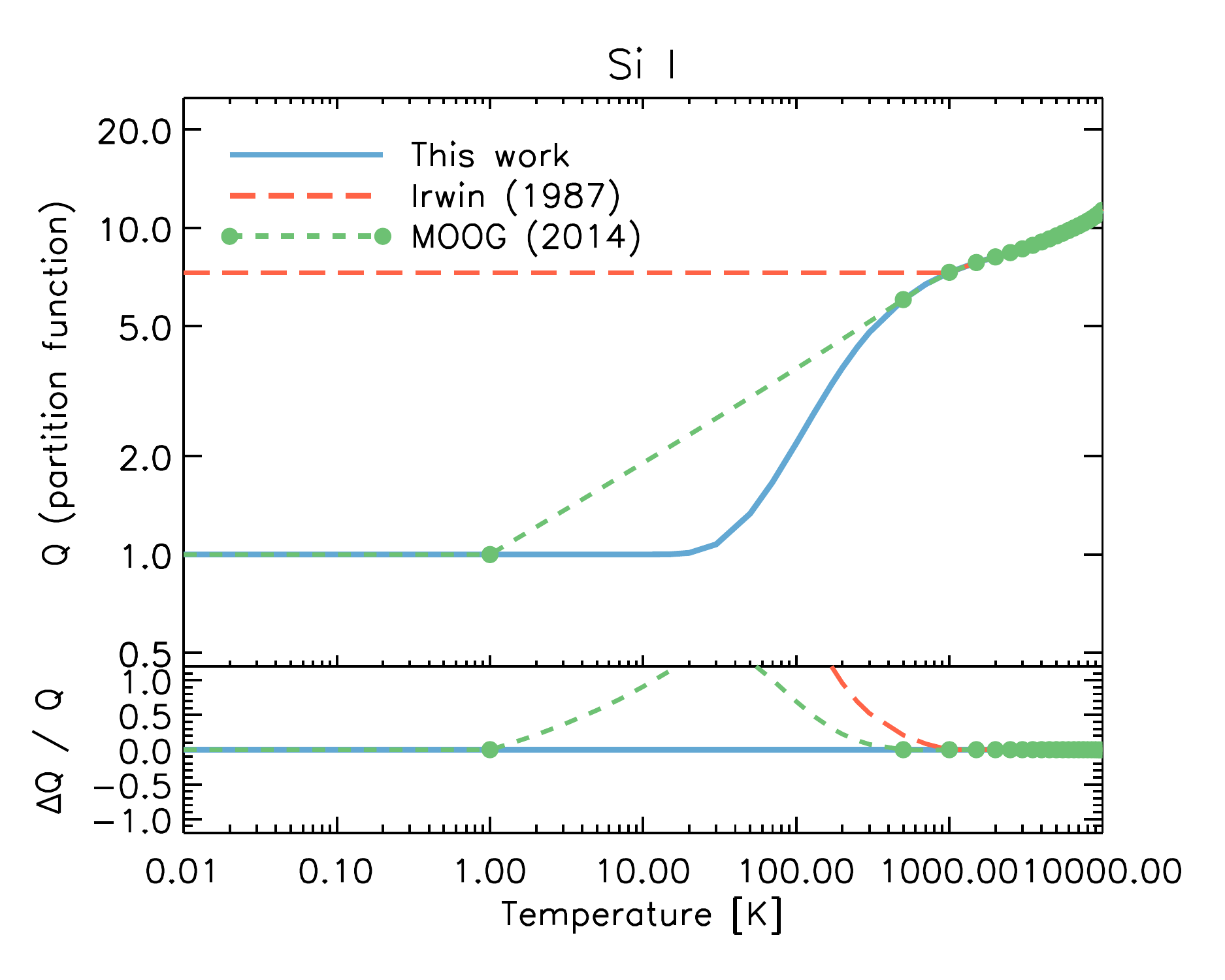}
}
\resizebox{0.9\hsize}{!}{
	\includegraphics{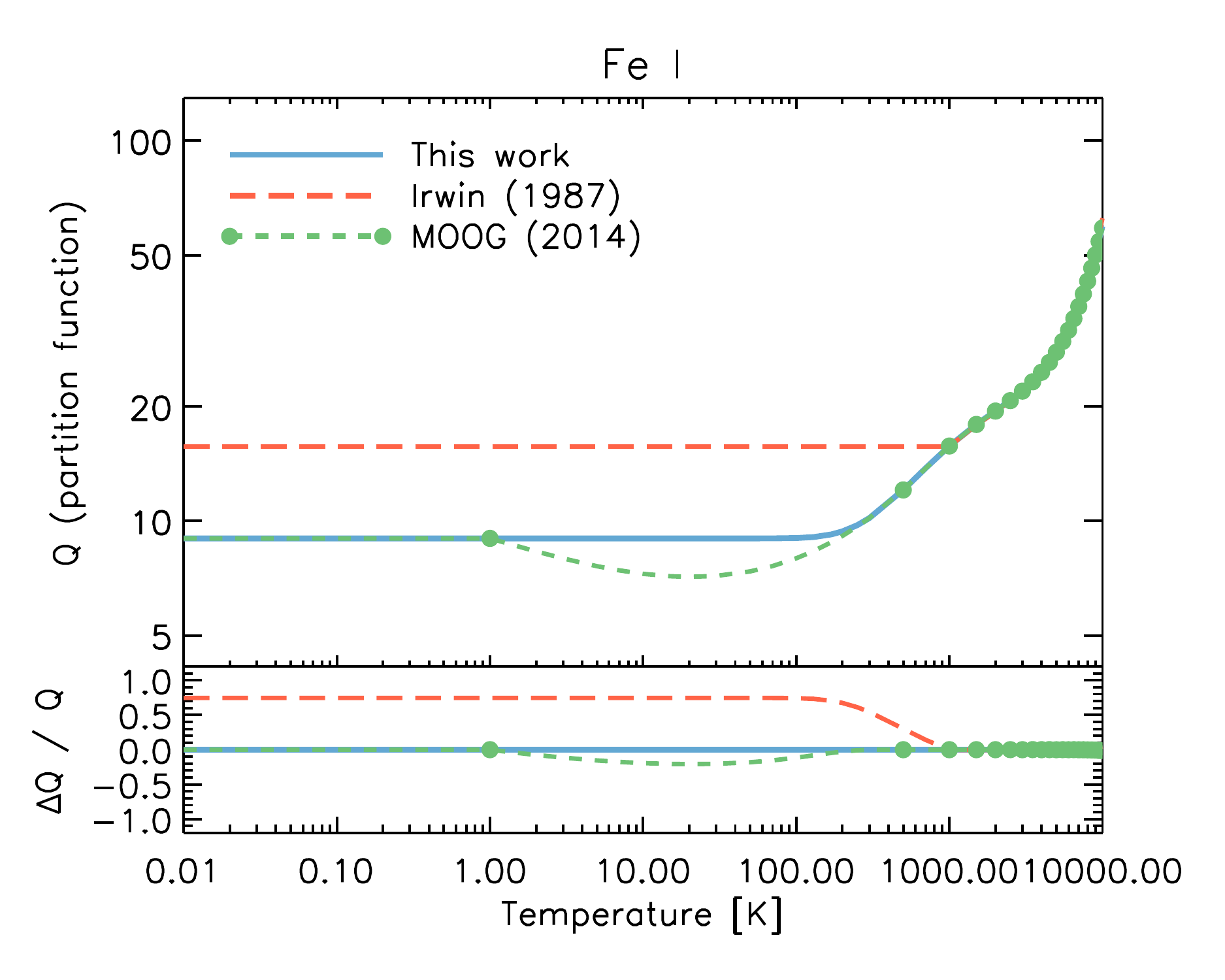}
	\includegraphics{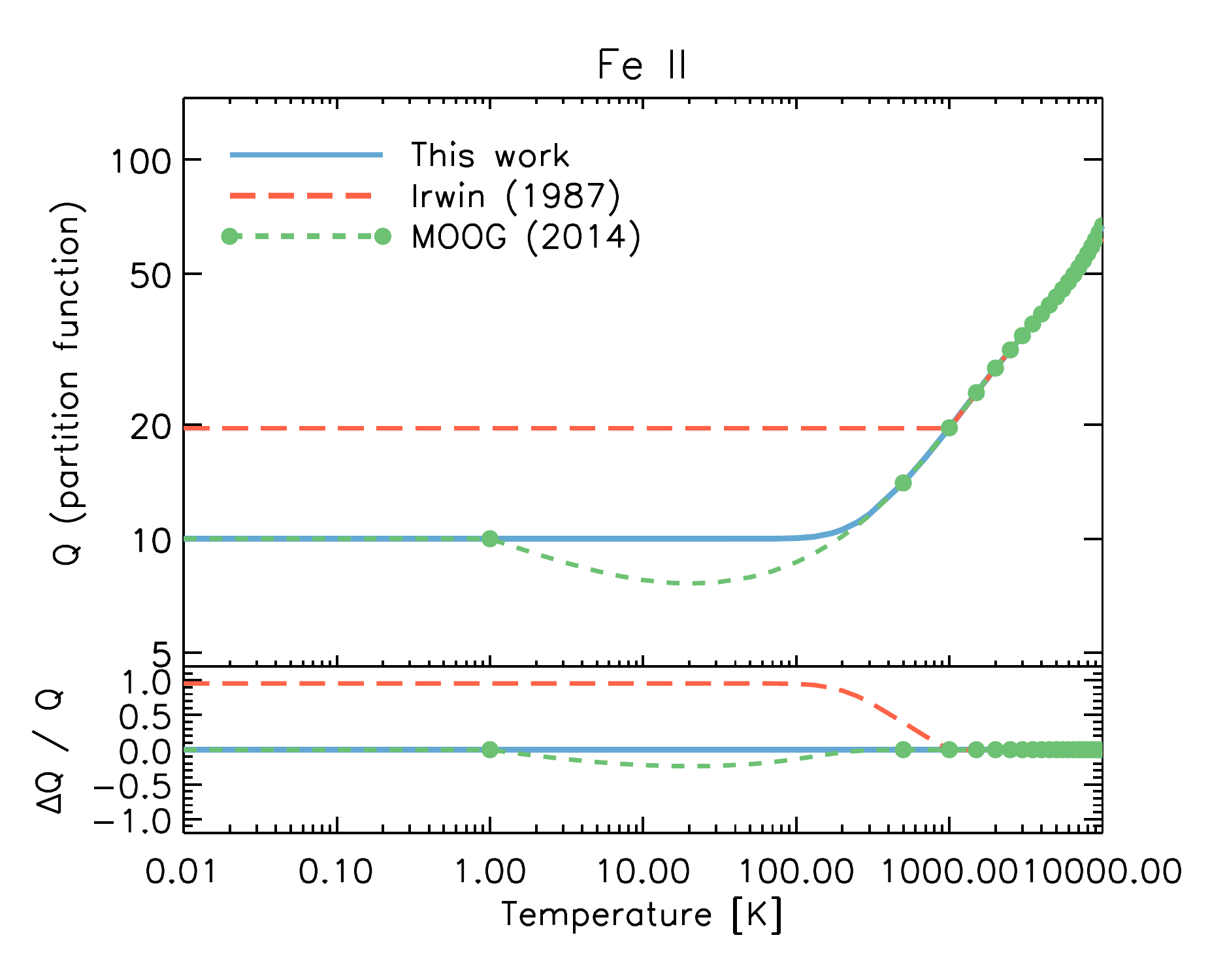}
}
\caption{\emph{Upper panels}: Comparison between the atomic partition functions (Q) calculated in this work with data from NIST (\emph{blue curves}) and the ones provided by \cite{1987A&A...182..348I} (\emph{red curves}) and by {\lawler} (MOOG data, version from July 2014; see also \citealt{Sneden:1973} and updates; \emph{green dots}). Partition functions from \cite{1987A&A...182..348I} are extended to temperatures below ${1\,000}$~K using  constant extrapolation (\emph{red dashed curves}). Data from {\lawler} are interpolated in between the sampled temperature points using cubic splines in $\log{Q}$. \emph{Lower panels}:  relative differences between the partition function calculations by the other authors and ours. }
\label{fig:atompf}
\end{figure*}

\begin{figure*}[htp]
\centering
\resizebox{0.9\hsize}{!}{
	\includegraphics{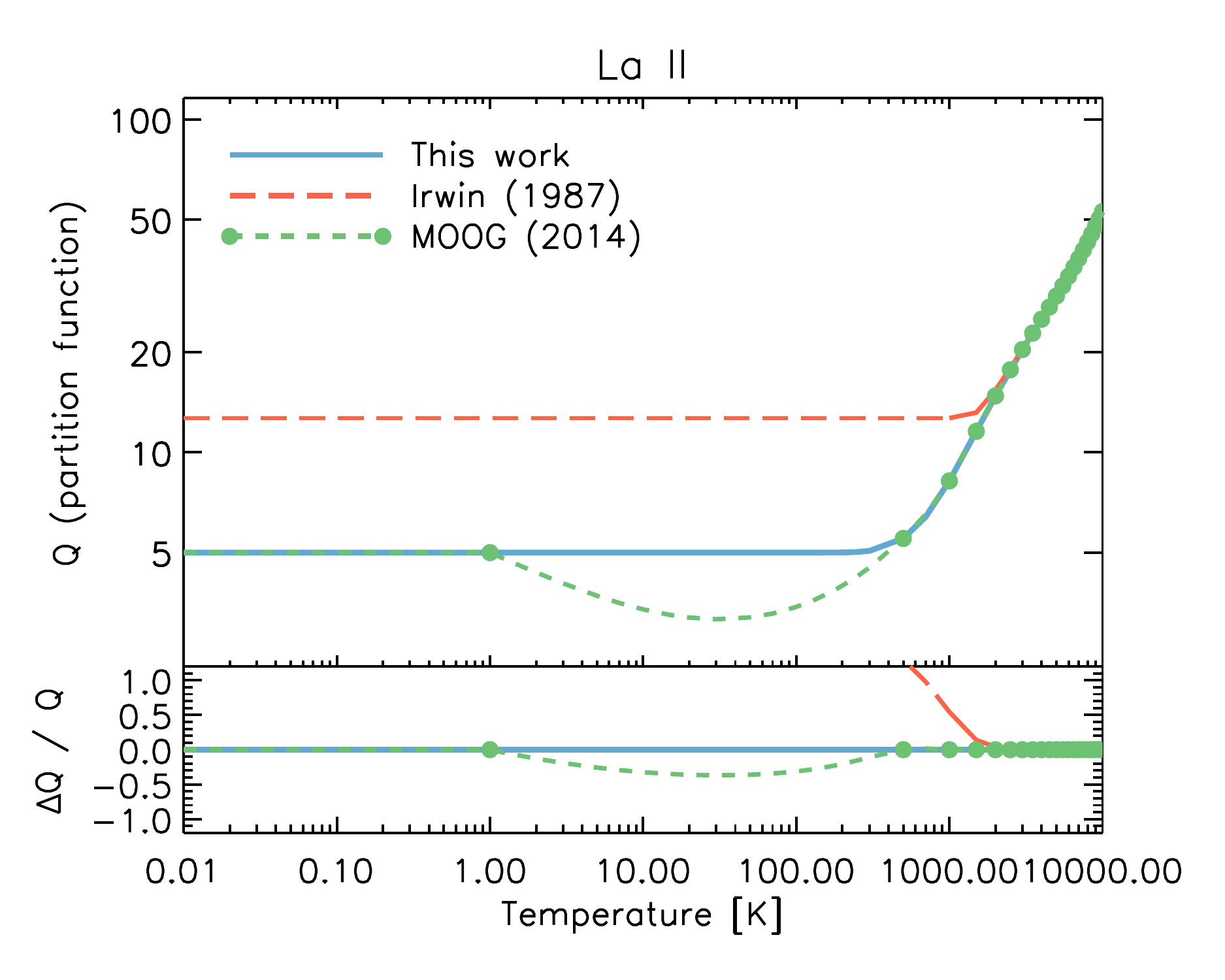}
	\includegraphics{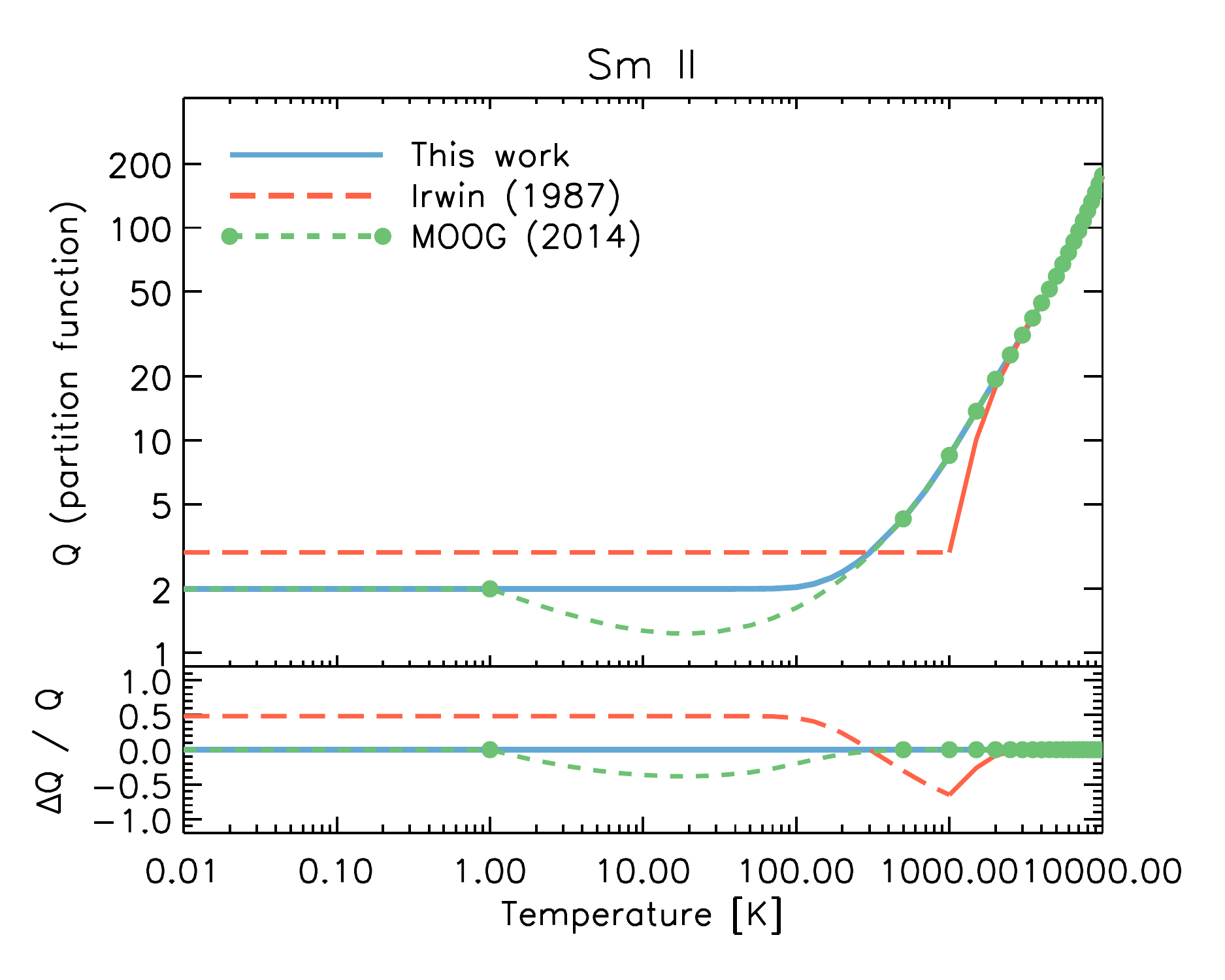}
}
\resizebox{0.9\hsize}{!}{
	\includegraphics{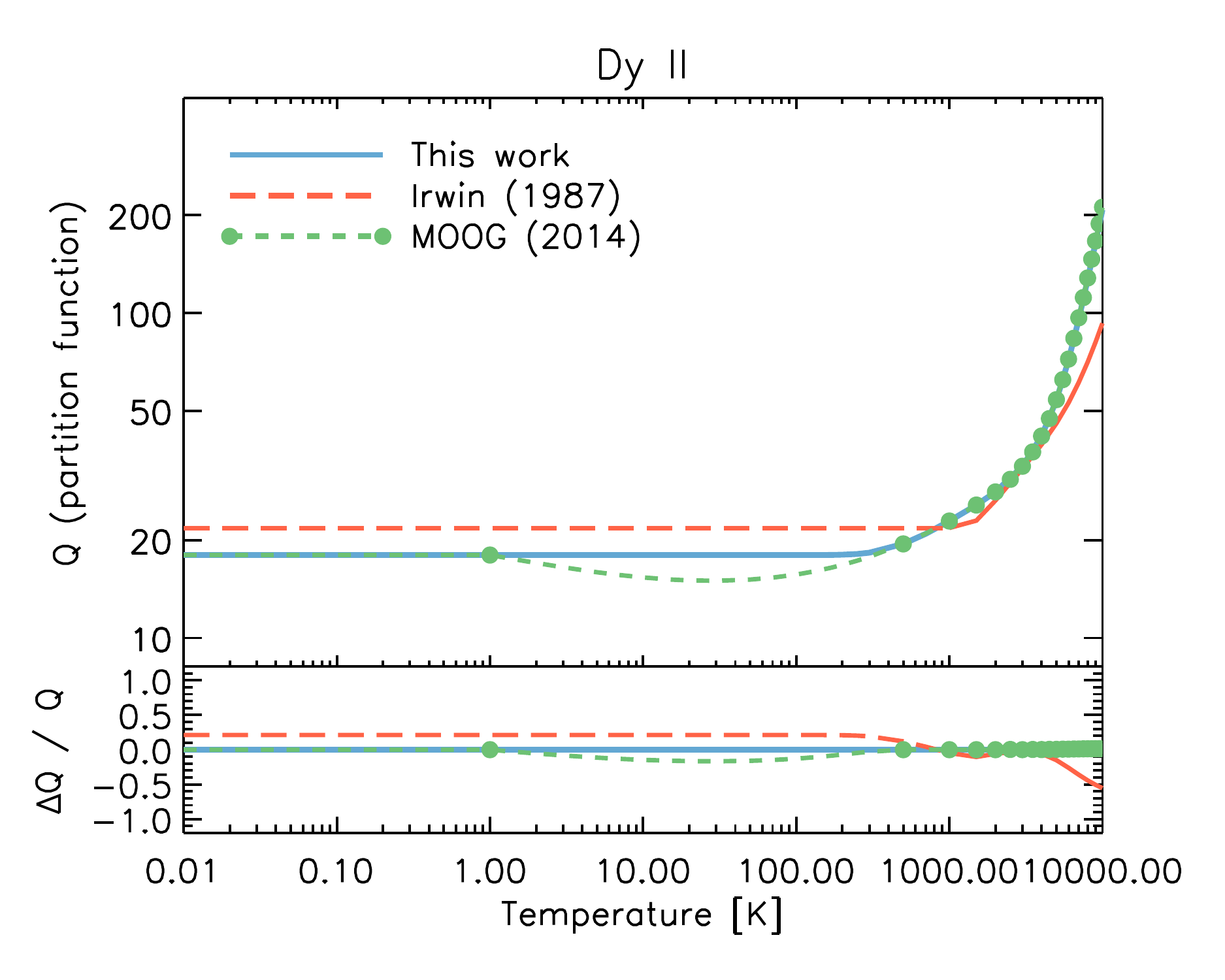}
	\includegraphics{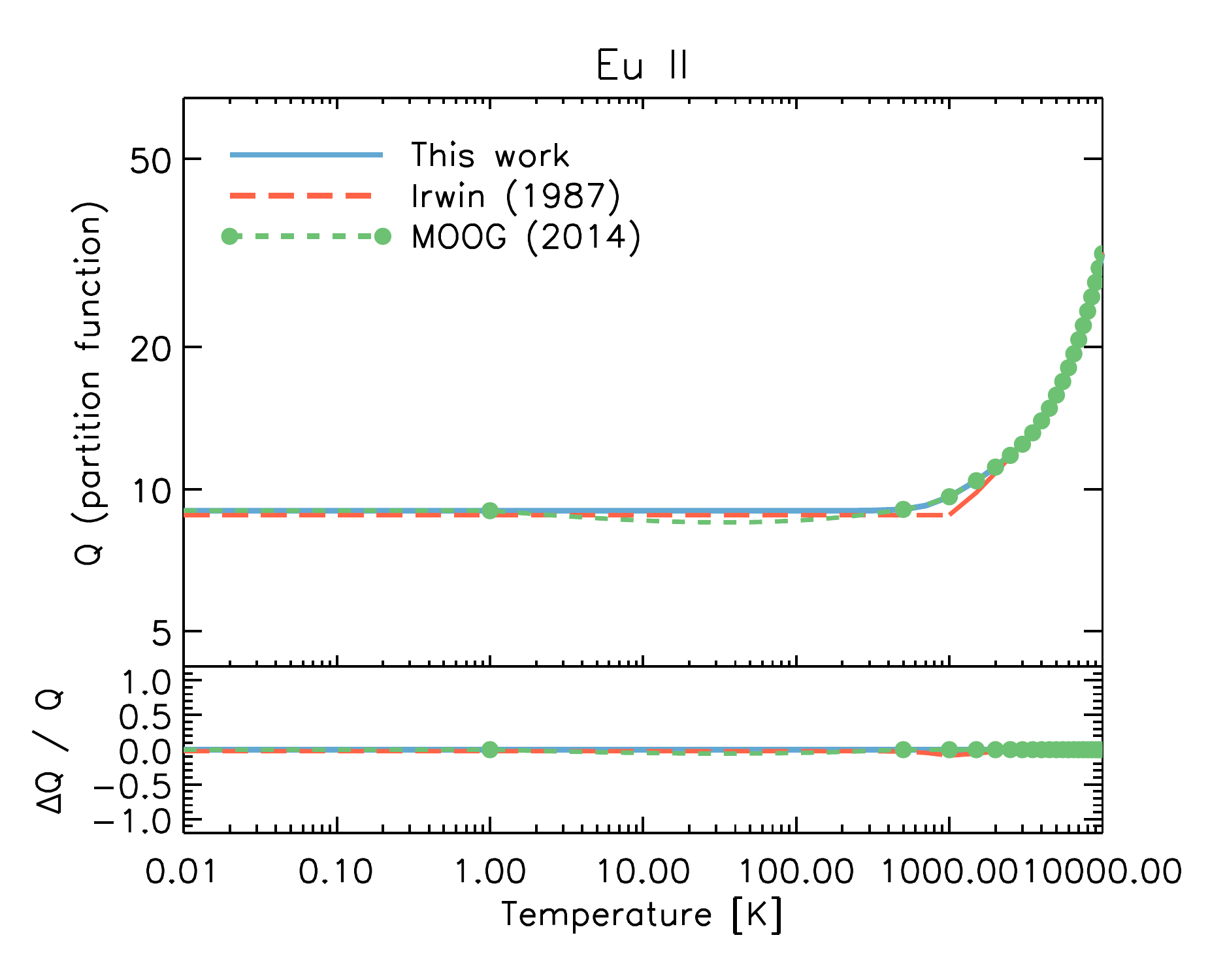}
}
\resizebox{0.9\hsize}{!}{
	\includegraphics{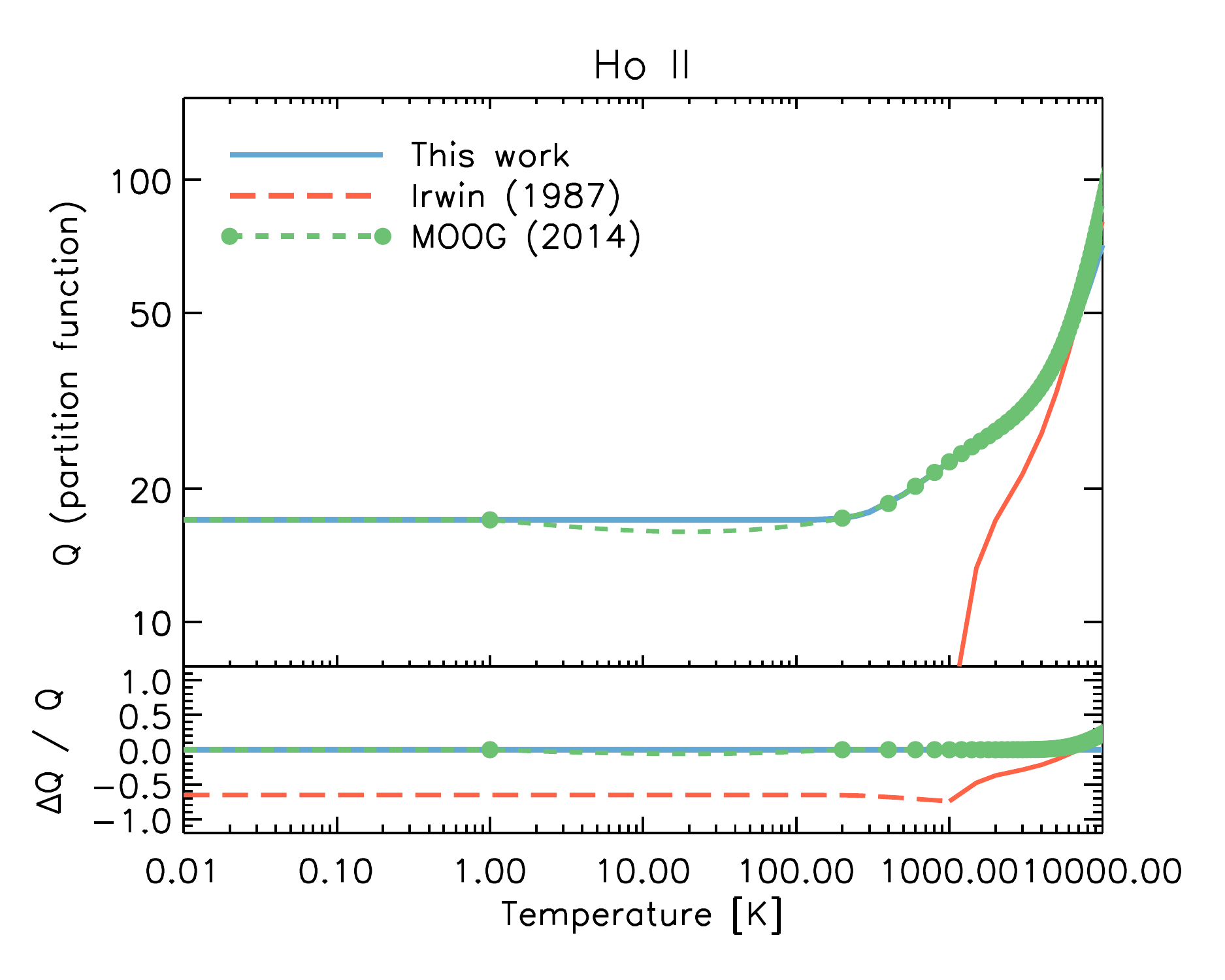}
	\includegraphics{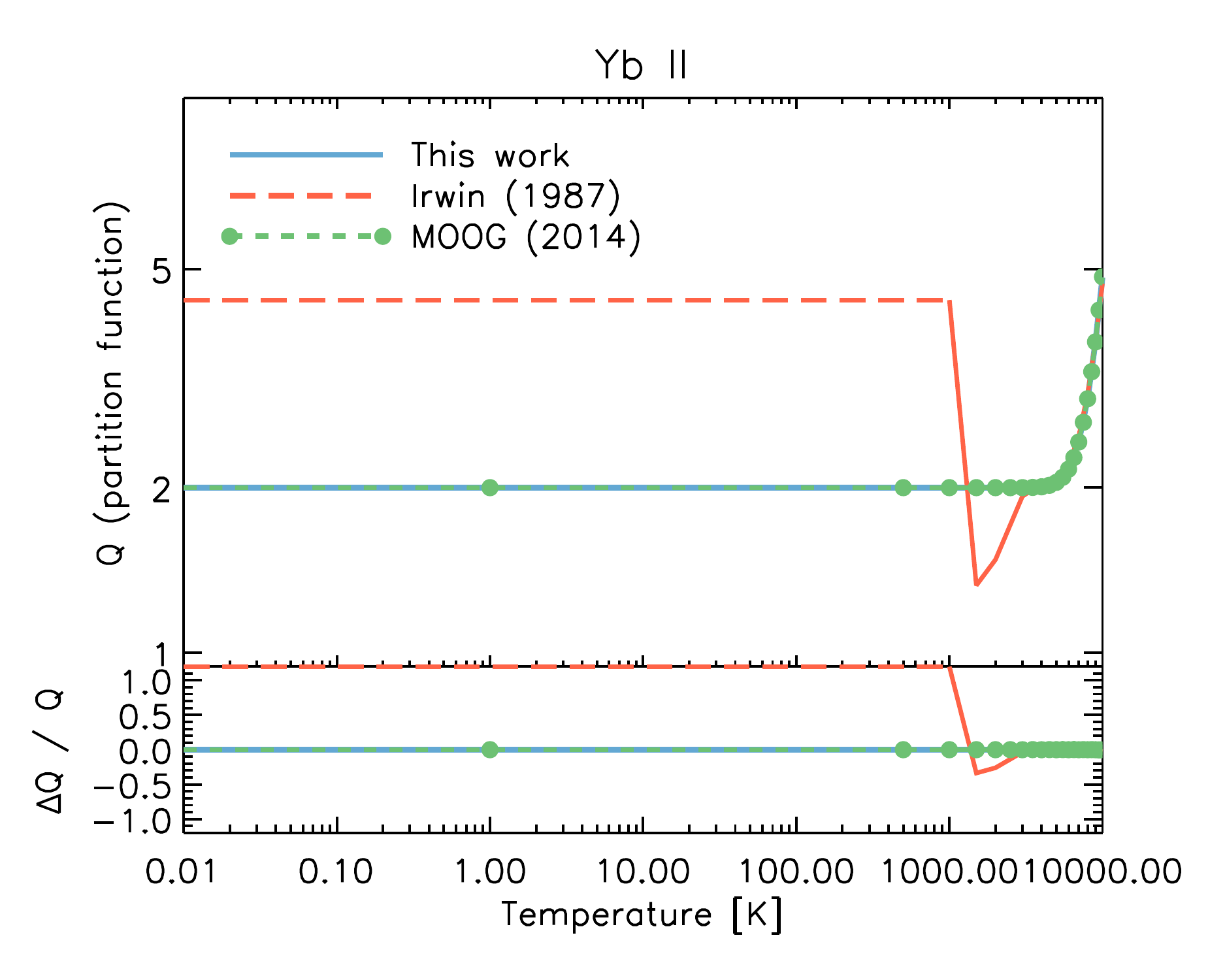}
}
\caption{Similar to Fig.~\ref{fig:atompf}, for some of the lanthanide ions. Note, the \ion{Yb}{ii} partition function from polynomial fits from \cite{1987A&A...182..348I} shows an unphysical sharp decrease between $1\,000$ and $2\,000$~K.}
\label{fig:atompf2}
\end{figure*}

\begin{figure*}[htp]
\centering
\resizebox{0.9\hsize}{!}{
	\includegraphics{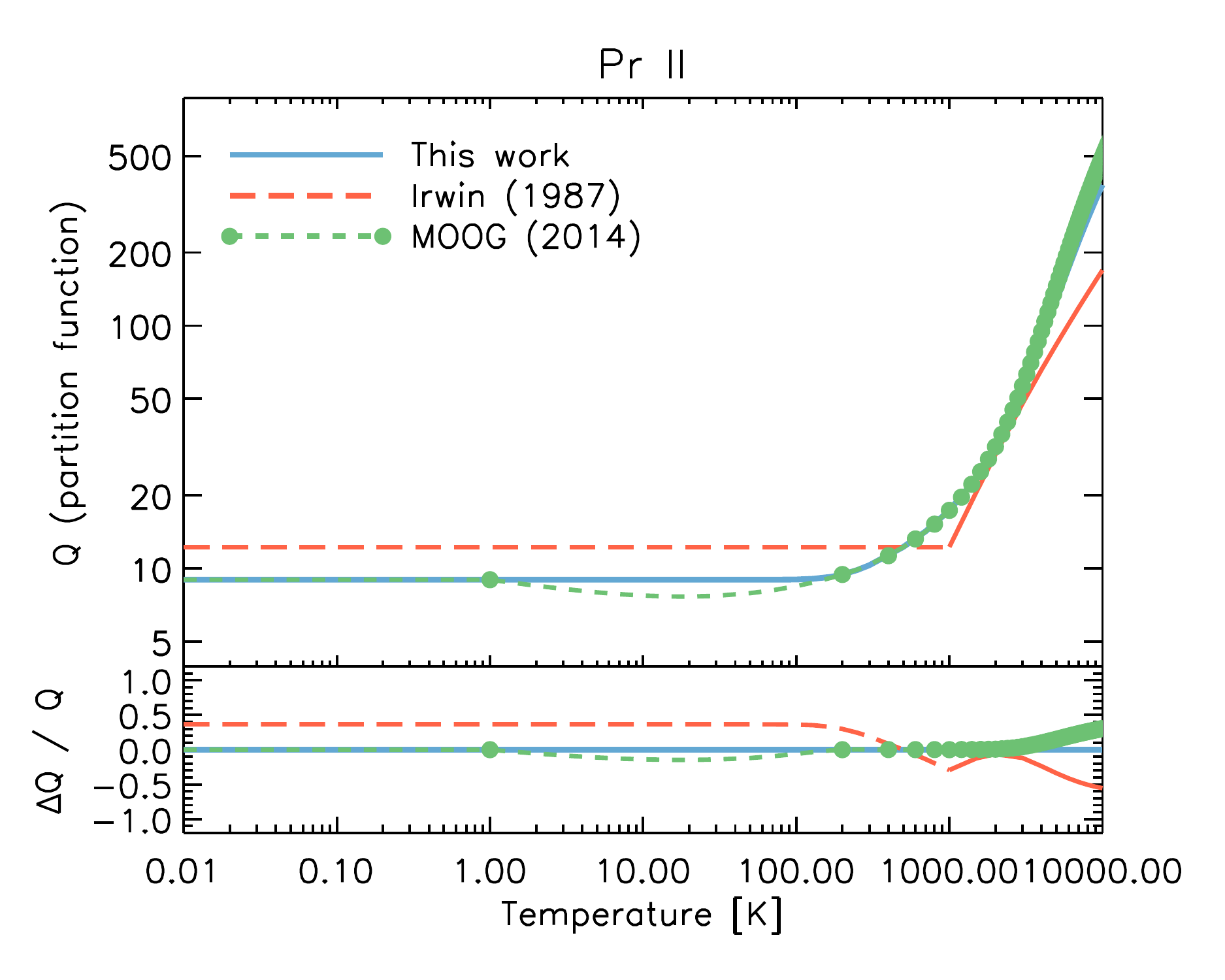}
	\includegraphics{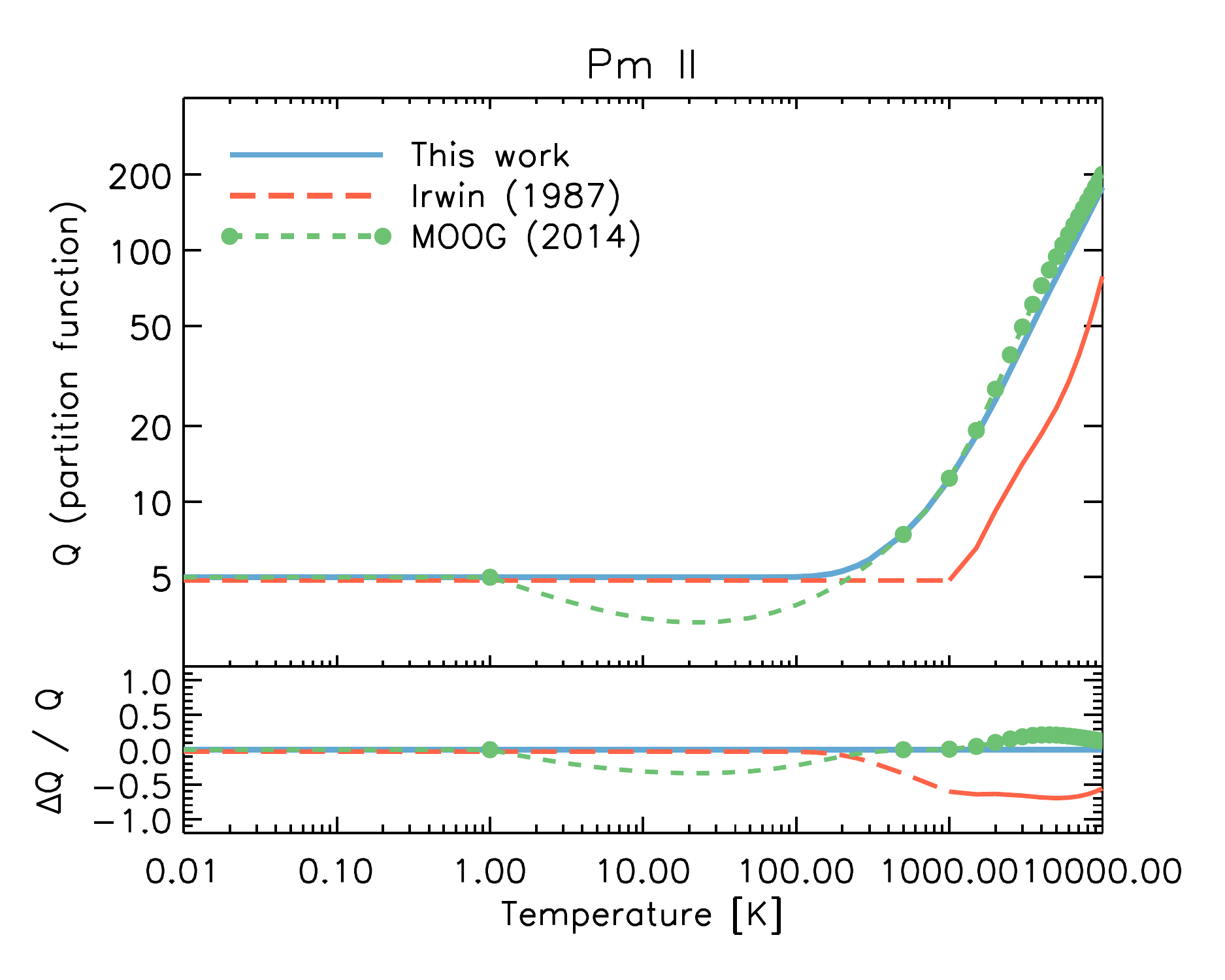}
}
\caption{Similar to Fig.~\ref{fig:atompf} and \ref{fig:atompf2}, for \ion{Pr}{ii} and \ion{Pm}{ii}, which, among the lanthanide ions, show the largest differences in terms of partition function values between our NIST-based calculations (\emph{blue curves}) and {\lawler}'s (\emph{green dots}) data at temperatures below $10,000$~K.}
\label{fig:atompf3}
\end{figure*}

  \begin{figure}[ht]
   \centering
   \includegraphics[width=90mm,angle=0]{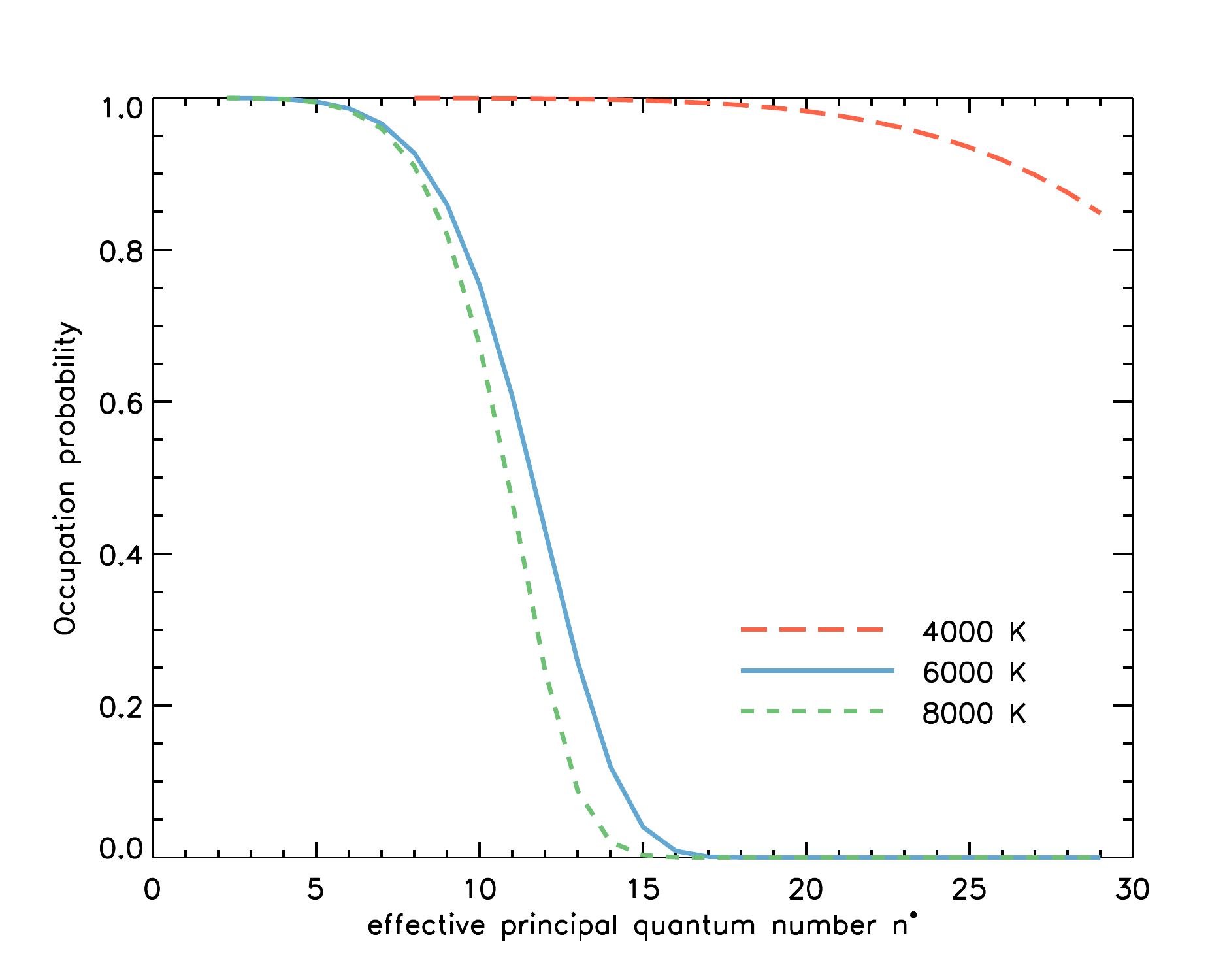}
      \caption{Plot of the occupation probability as a function of effective principal quantum number $n^*$ for conditions representative of three depths in the solar atmosphere.  The depths are labelled by their approximate temperatures, and these correspond to particle number densities of $N_\mathrm{H}$ and $N_\mathrm{e}$, of $10^{14}$ and $10^{13}$~cm$^{-3}$ at $4\,000$~K, of $10^{17}$ and $3 \times 10^{13}$~cm$^{-3}$ at $6\,000$~K, and of $10^{17}$ and $7 \times 10^{14}$~cm$^{-3}$ at $8\,000$~K, respectively.  In each case,   $N_\mathrm{He} = 0.1 \times N_\mathrm{H}$.}
         \label{fig:wplot}
   \end{figure}

  \begin{figure}[ht]
   \centering
   \includegraphics[width=90mm,angle=0]{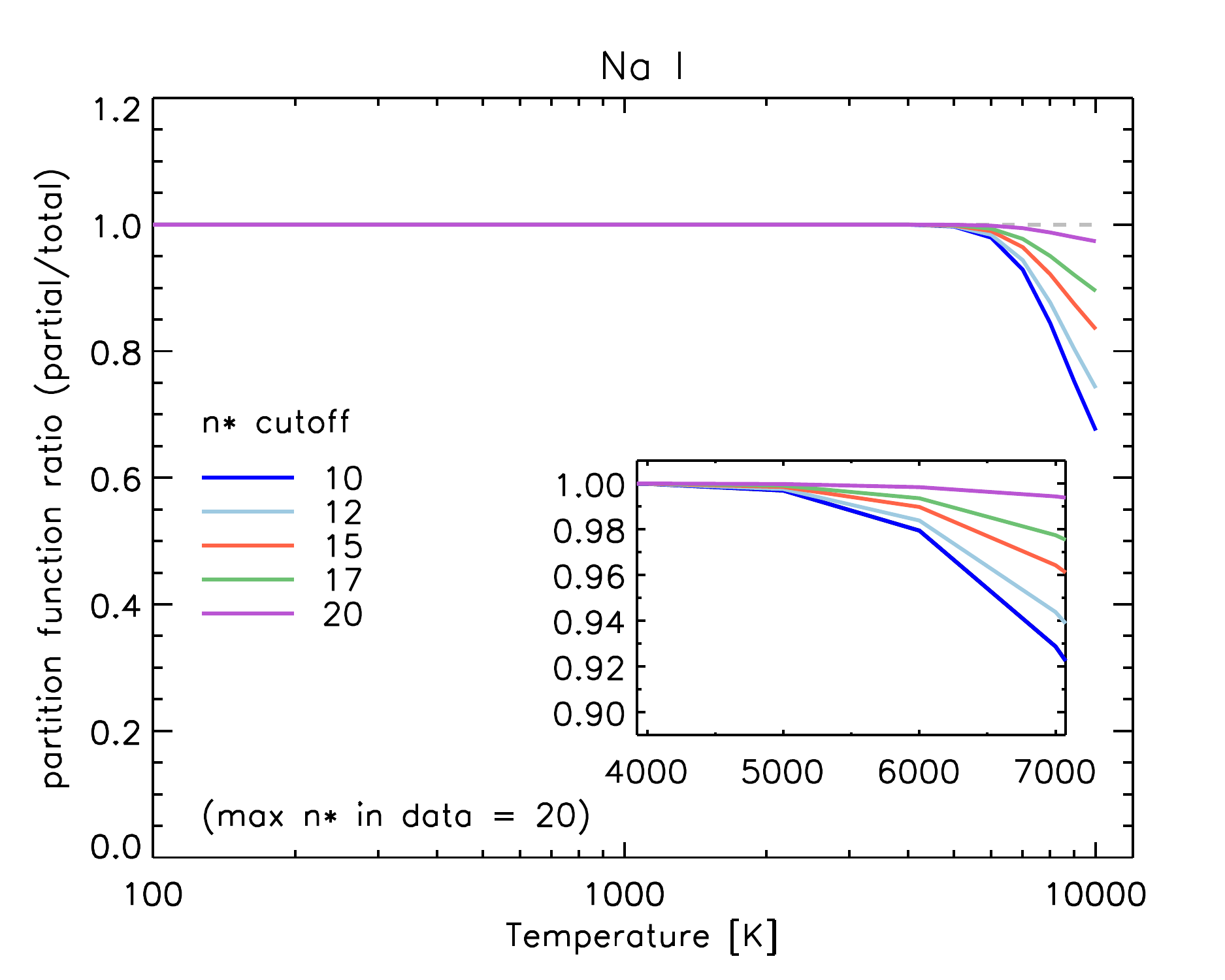}
   \includegraphics[width=90mm,angle=0]{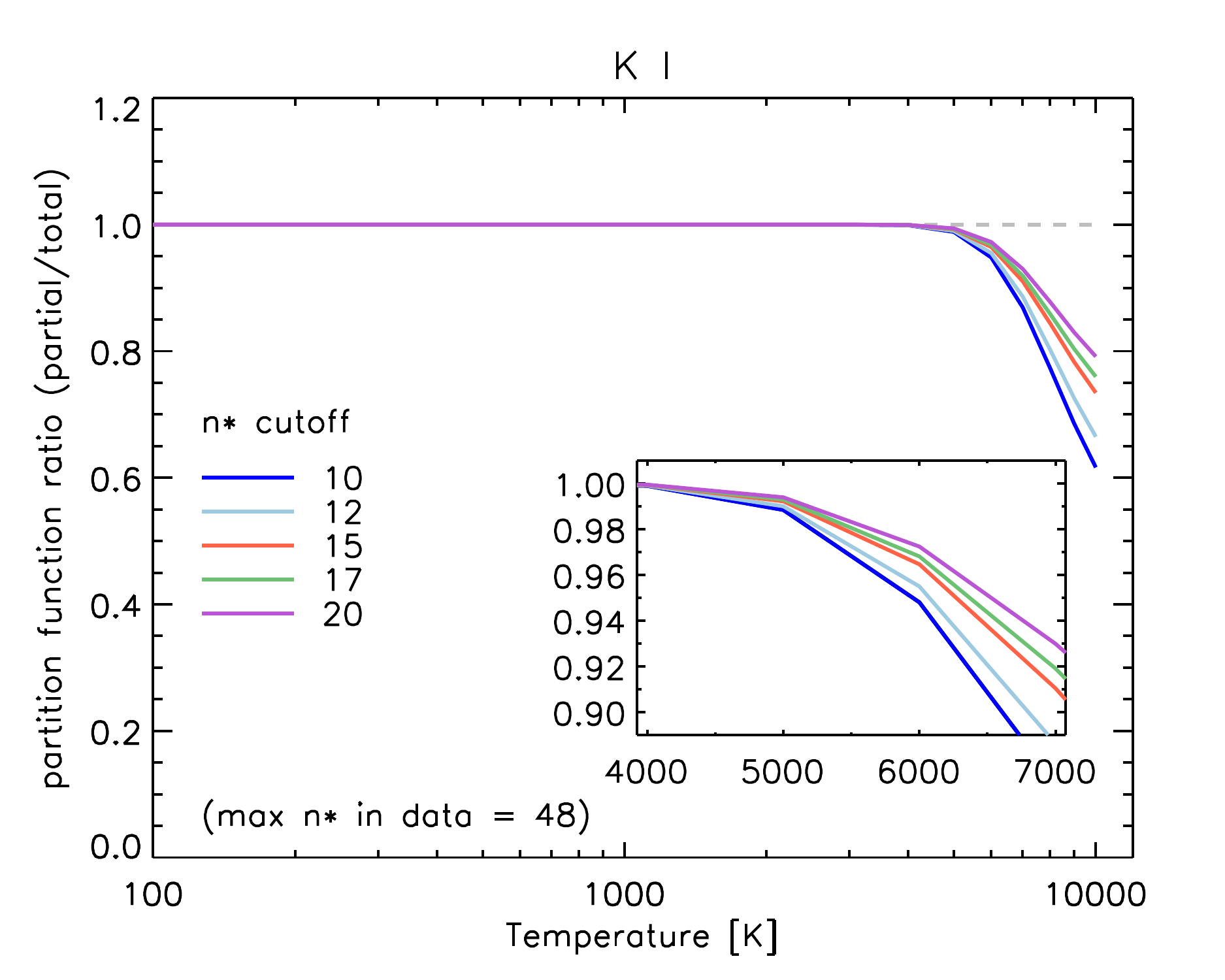}
      \caption{Plot of the ratio of partial partition functions calculated with a $n^*$ cutoff in the summation between 10 and 20, to the partition function calculated in this work including all states from NIST.}
         \label{fig:cutoff}
   \end{figure}

Our partition functions for atoms and ions are overall in very good agreement with results from \cite{1987A&A...182..348I} and {\lawler} (private communication), where the temperature ranges considered by the various sets of calculations overlap.  The {\lawler} data  are the basis of the partition functions used in the {\sc MOOG} spectral synthesis code (\citealt{Sneden:1973} and updates).  The original partition function data, kindly provided to us by {\lawler} are in some cases also based on NIST data, e.g. \ion{Tb}{ii} \citep{lawler_atomic_2001} and \ion{Ce}{ii} \citep{lawler_improved_2009}, while for a few cases taken from other sources, e.g. \ion{Ho}{ii} taken from \cite{bord_abundance_2002} \citep[see][]{lawler_improved_2004}.  These authors, however, provide polynomial fits or partition function data that are optimised or tabulated only for temperatures above ${\sim}{500}-{1\,000}$~K and below ${\sim}{10\,000}-{16\,000}$~K but are generally not applicable or available outside the designated range. For instance, the values of the polynomial fits provided by \cite{1987A&A...182..348I} would diverge very rapidly for temperatures outside the indicated range and are therefore not suited for applications to low-temperature astrophysical environments. {\lawler} do provide data at 1~K, but the ${1}-{500}$~K range is generally not sampled (or is, at best, largely undersampled), which makes it difficult in practice to accurately interpolate the values of the partition functions for in-between temperatures.
In Fig.~\ref{fig:atompf}, we present some examples of partition functions we calculated for various atoms and ions of astrophysical importance for molecule formation and stellar spectroscopy and compare our results with the ones from \cite{1987A&A...182..348I} and {\lawler}.  Agreement is generally good where the respective temperature ranges of the various sources overlap ($T>{1\,000}$~K), but information at lower temperatures are missing in the \cite{1987A&A...182..348I} and {\lawler} data sets. Simple extrapolation or, where possible, interpolation thereof yields inaccurate results at low temperatures.
Fig.~\ref{fig:atompf2} shows an analogous comparison for some lanthanide ions of astrophysical interest for stellar spectroscopy of neutron-capture elements. Agreement among the various sources is less satisfactory here, with the \cite{1987A&A...182..348I} data set showing the largest deviations with respect to our calculations and {\lawler}'s. The last two are in excellent agreement with each other in this case, the only limitation of the {\lawler} data set being again the sparser sampling at low temperatures.  The largest differences are found for \ion{Pm}{ii} and \ion{Pr}{ii} (Fig.~\ref{fig:atompf3}) for which our partition functions at high temperatures are ${\sim}20$-$30$\% lower than the {\lawler} ones.

As mentioned in Sect.~\ref{sect:calc}, the summation over states in the partition function calculations for atoms and molecules is performed over all energy levels for which data are available.  In the case of molecules, only a few cases such as H$_2$ have a significant number of identified highly-excited states.  As seen in Fig.~\ref{fig:molpart} the exclusion of states above $40\,000$~cm$^{-1}$ has practically no effect on the partition function.  The most likely scenario for molecules is that missing high-lying states leads to an underestimate of the partition function, leading to an overestimate of the equilibrium constant.  For cases such as H$_2$ this error can be expected to be small at all temperatures, but in cases where few excited states are identified this may lead to significant errors particularly at higher temperatures.   

In the case of atoms, NIST often provides a very large number of Rydberg states.  \cite{hummer_equation_1988a} investigated the sensitivity of a hydrogenic partition function to truncation of the summation at a given cutoff effective principal quantum number $n^*$.  They showed that sensitivity to the choice of cutoff is dependent on $I/kT$, where $I$ is the ionisation potential of the species of interest.  In particular, when $I/kT$ is relatively large, say greater than 10, the ground state dominates the partition function, and the result is quite insensitive for any reasonable choice of the cutoff.  However, as $I/kT$ becomes smaller the partition function becomes sensitive to the cutoff.  Thus, at any given temperature we expect species with low ionisation potentials (i.e. neutral species of alkalis and alkaline earths) to be the most sensitive to the cutoff procedure, and for any given species the sensitivity increases with temperature.  

To investigate this we performed calculations in the chemical picture using the occupation probability formalism \citep{hummer_equation_1988a}.  The occupation probability defines the likelihood that a given state is not dissolved by the perturbations of nearby particles.  We calculate this probability for conditions approximating three depths in the solar atmosphere, using eqn. 4.71 of \cite{hummer_equation_1988a}, where the part relating to perturbations by charged particles is replaced with the expressions from Appendix A of \cite{hubeny_nlte_1994}.  The results are shown in Fig.~\ref{fig:wplot}, and indicate that in the typical line forming regions of a solar-type stellar atmosphere, states with $n^*\ga 12$ will be dissolved and these states should be excluded from the partition function summation.  We performed calculations for a significant number of elements where a cutoff was implemented for values between $n^*=10$ and $20$.  For neutral species of many atoms of astrophysical interest, such as C, O, Mg, and Fe, and for all ionised species, the results showed no significant changes compared to the final results including all states across the entire temperature range; much less than the per cent level.  However, as expected, neutral species of alkalis, which  have low ionisation potentials, showed some effects at higher temperatures; see Fig.~\ref{fig:cutoff}.  At temperatures corresponding to the line forming regions of solar-type stars, around $5\,000$--$6\,000$~K, which indeed form lines of neutral alkalis, we see that differences for a cutoff of $n^*\approx 12$ are of order a couple of per cent.  This indicates that in such cases, the partition function calculated here would be overestimated, and if greater accuracy is required, the plasma conditions must be accounted for.  We note that the effects are even larger at higher temperatures, but very few neutral atoms will be found at these temperatures, except at high densities where plasma interactions must be accounted for.  In general these results are not appropriate for environments with densities much greater than those typical of stellar atmospheres.

\section{Concluding remarks}

Calculations of partition functions and equilibrium constants for 291 diatomic molecules have been presented, and dissociation energy data collated for these molecules.  The calculations are based on spectroscopic constants from HH, with updates to constants for the ground states of 85 molecules by \cite{2007JPCRD..36..389I, 2009JPCRD..38..749I}, covering the most important molecules.  We note that there are other compilations containing partition function data for some diatomic molecules.  For example \cite{2000JMoSt.517..407G} for CO, O$_2$, NO, OH, HF, HCl, HBr, HI, ClO, N$_2$, and NO$^+$, over the temperature range 70 to 3005~K, and \cite{2009IJT....30..416B} for N$_2$, N$_2^+$, NO, O$_2$, CN, C$_2$, CO, and CO$^+$, up to $50\,000$~K, though these data are only available on request.  We note that all but two of these molecules are covered by the updates from Irikura; the exceptions being HBr, which is not covered by our study, and HI, which is probably of limited astrophysical importance.  Thus, we expect that the calculations presented here for these cases are of similar accuracies to these studies.  For example, in the important case of CO, we note that the spectroscopic constants employed by us and those employed by \cite{2009IJT....30..416B} for the ground state agree in all cases to better than four significant figures, and it was seen in Fig.~\ref{fig:molpart} that the partition functions agree almost exactly.  We also note that there are numerous single molecule studies, e.g. \citet{1996JQSRT..55..849S,2005MNRAS.357..471E}, often based on energy levels derived from explicit solution of the Schr\"odinger equation in quantum chemistry potentials.  Merging these results into our results would be non-trivial, primarily due to differences in the temperature ranges.  We emphasise that if a specific molecule of interest is important it may be worth to compare the results from these calculations with those from single molecule studies, and perhaps replace the data.

Partition functions have also been calculated for 284 species of atoms for all elements from H to U.  The data are based on up-to-date NIST critically compiled data.  We have investigated the possible influence of inclusion of many Rydberg states, and estimate that this should not lead to uncertainties larger than a couple of per cent in typical applications.  If higher accuracy is required, then more sophisticated methods allowing for interactions with the environment may be required.  

Finally, we comment that critical compilation of partition functions and/or the data necessary to calculate them has become a task that is probably beyond any individual or even small group.  It seems to us that the most reasonable approach is a distributed one, where the community builds a definitive set of partition functions and spectroscopic constants and/or potentials and/or energy levels, updated via gradual improvements from studies of individual or small sets of molecules.  This would require some kind of centralised infrastructure in the form of a database or similar, and would be important for both the astrophysics and molecular physics communities.

\begin{acknowledgements}
We are grateful to Jim Lawler and Chris Sneden for supplying their raw partition function data for atoms and ions.  We thank Nikolai Piskunov for stimulation of, and interest in, this project and support for visits by R.C.\ to Uppsala.  We are thankful to Alexander Kramida from NIST for assisting us with the compilation of the list of sources for atomic energy level data. We gratefully acknowledge the support of the Royal Swedish Academy of Sciences,  G{\"o}ran Gustafssons Stiftelse and the Swedish Research Council.  P.S.B.\ was a Royal Swedish Academy of Sciences Research Fellow supported by a grant from the Knut and Alice Wallenberg Foundation during much of this work.  P.S.B.\ is also supported by the project grant “The New Milky” from the Knut and Alice Wallenberg foundation.  R.C.\ acknowledges support from the Australian Research Council through a Discovery Early Career Researcher Award (DECRA) grant (project DE120102940).  
\end{acknowledgements}

\bibliographystyle{aa} 
\bibliography{equil}

\end{document}